\documentclass[amsmath,amssymb,11pt]{article}
\usepackage{jcappub}
\pdfoutput=1
\usepackage{epsfig}
\usepackage{float}
\usepackage{subfig}
\usepackage{subfloat}
\usepackage{graphicx}
\usepackage[utf8]{inputenc}
\usepackage{hyperref}
\usepackage{caption}
\usepackage{slashed}
\usepackage{tikz}
\usepackage[makeroom]{cancel}
\usepackage[hang,flushmargin]{footmisc}

\usepackage{color}

\global\long\def\g{\gamma}

\newcommand*{\affmark}[1][*]{\textsuperscript{#1}}

\newcommand{\beq}{\begin{equation}}

\newcommand{\eeq}{\end{equation}}

\title{Detecting the $\pi$-axiverse through parametric resonance}

\author{Stephon Alexander,\affmark[1]}
\emailAdd{stephon\_alexander@brown.edu}
\author{Geoff Beck,\affmark[2]}
\emailAdd{geoffrey.beck@wits.ac.za}
\author{Santiago Loane\affmark[1,3]}
\emailAdd{loane2@illinois.edu}
\author{and Tucker Manton,\affmark[1,4]}
\emailAdd{tucker\_manton@ucas.ac.cn}

\affiliation{\affmark[1]Department of Physics,
Brown University, Providence, RI 02912, USA\\
\affmark[2]School of Physics and Centre for Astrophysics, University of the Witwatersrand, Johannesburg, 1 Jan Smuts Ave, WITS-2050, South Africa\\
\affmark[3]Illinois Center for Advanced Studies of the Universe, Department of Physics,
University of Illinois Urbana-Champaign, Urbana, IL 61801, USA \\
\affmark[4] School of Fundamental Physics and Mathematical Sciences, Hangzhou Institute for Advanced Study, University of Chinese Academy of Sciences (HIAS-UCAS), Hangzhou, 310024, China.
}
\abstract{Axions are a leading dark matter candidate. In this work, we study the detectability of a multi-axion-like model, dubbed the $\pi$-axiverse,  that is distinguishable from the string axiverse. The dark matter candidates are $N^2-1$ pseudo-Nambu-Goto modes (pion- and kaon-like states) stemming from spontaneous breaking of a global $SU(N)$ flavor symmetry. The low energy theory includes $N-1$ axionic couplings with additional couplings to the Standard Model photon kinetic energy, 
reminiscent of the string theory dilaton-photon coupling. We explore the parametric resonance of photons interacting with such a dark sector. Axions are well known to form macroscopic solitonic-like objects (axion stars), which experience instabilities due to overdensities stemming from mergers or accretion processes. The instabilities produce high-intensity bursts of radiation via parametric resonance that may be detected at observatories such as MeerKAT, the Square Kilometre Array (SKA), and the next generation Very Large Array (ngVLA). Using numerical methods, we systematically explore the multi-dimensional parameter space of the $\pi$-axiverse to search for regions where such signals are detectable, which generically differ from single axion models. We identify regions of the parameter space where MeerKAT, SKA, and ngVLA can resolve such signals, assessing the potential of transient searches to constrain the model. Our results provide a significant step forward in understanding the phenomenology and indirect detection of multi-axion-dilaton dark matter. 

}

\begin{document}

\maketitle

\section{Introduction} \label{sec:intro}

Dark matter constitutes about one-fourth of the energy budget of the universe, yet its fundamental nature remains one of the most pressing mysteries in modern physics. With the WIMP parameter space slowly being ruled out by experiments on all fronts, axions have taken center stage as a leading dark matter candidate. Initially proposed by Peccei and Quinn to resolve the strong CP problem in quantum chromodynamics (QCD) \cite{Peccei:1977hh}, work by Weinberg \cite{Weinberg:1977ma} and Wilczek \cite{Wilczek:1977pj} further showed that the axion has properties aligning naturally with cold dark matter. Axions and axion-like particles (ALPs) arise as a cosmological relic through various production mechanisms such as misalignment \cite{Preskill:1982cy,Marsh:2015xka,Sikivie:2006ni} (see also \cite{DiLuzio:2020wdo} for a review of QCD axion models). 

ALPs are also a generic prediction in string theory, appearing as zero-modes of antisymmetric tensors during compactification \cite{Svrcek:2006yi}. Although stringy axions generically couple to photons \cite{Agrawal:2024ejr}, the number of stringy ALPs in the low-energy four-dimensional theory can be extremely large with a logarithmically hierarchical mass spectrum \cite{Cicoli:2012sz}. This is famously dubbed `the String Axiverse' \cite{Arvanitaki:2009fg}. In \cite{Alexander:2023wgk}, two of the current authors proposed a multi-axion model that is an alternative to the string axiverse, dubbed the `$\pi$-axiverse.' The model expands on ideas initiated in \cite{Maleknejad:2022gyf}, and was later generalized in \cite{Alexander:2024nvi}. The $\pi$-axiverse stems from a dark copy of the Standard Model with ultralight quarks, and the relic dark matter candidates are analogous to the Standard Model pions and kaons. Specifically, for a $SU(N_f)$ flavor symmetry, we have $N_f^2-1$ $\pi$-axions that may be real, complex and electrically charged, or complex neutral scalars. In addition to the standard three-point (neutral) pion coupling to the photon $\pi_0 F\tilde{F}$, identical to the axion coupling, the $\pi$-axiverse predicts four-point couplings to the photon kinetic energy. These are similar to string theory dilaton couplings, therefore the $\pi$-axiverse is reminiscent of a multi-axion-dilaton model. In difference to the string axiverse, the masses of the $\pi$-axions lie in a tightly spaced spectrum, with separations of $O(10^0-10^2)$. The model parameters are such that the $\pi$-axions can naturally have masses in the $\mu$eV range, therefore we can leverage ongoing detection efforts which probe precisely that mass window. We will show in this work that multi-axion models of this type have significantly different phenomenology than better studied, single axion models, with unique signatures and exciting detection prospects. 

The feeble axion couplings to Standard Model particles make detection extraordinarily challenging, yet their predicted two-photon coupling  enables electromagnetic searches. Resonance cavity detection was proposed decades ago \cite{Sikivie:1983ip}, wherein axions convert to photons in a strong magnetic field, a technique now employed by experiments such as the Axion Dark Matter Experiment (ADMX) \cite{Asztalos_2010,ADMX:2018gho}. Advances in quantum amplification and cavity design \cite{Sikivie:1983ip} have pushed sensitivity into the theoretically favored axion mass range, while alternative approaches (e.g., dielectric haloscopes \cite{Jaeckel:2010ni} and nuclear magnetic resonance \cite{Budker:2013hfa}) diversify the search landscape. Many different experimental searches are currently underway (see, \textit{e.g.} \cite{Antypas:2022asj} for a fairly recent review). 

While laboratory experiments target local axion dark matter, astrophysical observations provide complementary constraints through indirect signatures. Axions may convert to photons in the magnetized environments of neutron stars \cite{Hook:2018iia}, generating narrow radio signals detectable by telescopes like MeerKAT \cite{Zhou:2022yxp} and the future Square Kilometre Array (SKA)\footnote{In \cite{Zhang:2024pys}, the authors analyzed indirect DM detection prospects at SKA from DM annihilation for masses in the GeV - 100 GeV range, heavier than conventional ALP masses.} \cite{Braun:2015zta}. Surveys for unexplained radio lines from neutron stars (e.g., the ``X-ray-to-radio" correlation in magnetars \cite{Safdi:2018oeu}) or core-collapse Primakoff processes \cite{Payez:2014xsa} further probe axion-photon coupling. Axion clouds can also be formed around black holes through a superradiant instability \cite{Rosa:2017ury,Yoshino:2013ofa}, producing gravitational radiation and electromagnetic bursts. The synergy between high-sensitivity multimessenger astronomy and theoretical models is opening new avenues to explore axion parameter space beyond terrestrial experiments.  

Axions (and ultralight bosons in general \cite{Zhang:2024bjo}) form macroscopic, solitonic bound states broadly referred to as axion stars \cite{Eggemeier:2019jsu,Chen:2020cef,Widdicombe:2018oeo}. The dynamics of merging axion stars offer a unique window into axion-photon coupling, which is the focus of this paper. During mergers, the high-density axion field experiences instabilities, efficiently converting axions to photons via the $F\tilde{F}$ interaction \cite{Levkov:2016rkk}. This process may produce transient radio or X-ray bursts resembling fast radio bursts (FRBs) \cite{Iwazaki:2014wta, Iwazaki:2015zpb,Di:2023nnb,Di:2024jia}. Because the $F\tilde{F}$ interaction violates parity, the resulting photons are preferentially polarized. In fact, the analysis of FRB 20240114A revealed \cite{Xie:2024fgo} nontrivial polarization structure consistent with other FRB repeaters. Although such physics can be described by the presence of an axion (e.g. \cite{BetancourtKamenetskaia:2024gcv}), it is not a smoking gun; many astrophysical processes can polarize electromagnetic radiation.  

The conversion of axions to photons can take place through the process of \textit{parametric resonance}, where axion oscillations drive growth of the photon field amplitude. The growth happens for a finite period of time, while the system is above a critical threshold. The critical threshold can phenomenologically be reduced to a mass threshold, above which we see a burst of photons. We note that if the axion star mass exceeds the critical value for gravitational stability ($M \sim 10^{-12} M_\odot$ for the QCD axion \cite{Helfer:2016ljl}), collapses can trigger ``bosenova'' explosions, releasing a significant fraction of the mass energy as scalar radiation, i.e. free-streaming axions. The critical threshold for parametric resonance is not necessarily the same as for gravitational instability. In a merger event, one most likely has two lower density axion stars merging to form a denser star that can exceed either or both of the critical thresholds; see \cite{Hertzberg:2020dbk} for cases where only electromagnetic instability occurs after axion star merger, while \cite{Chung-Jukko:2024hod} explores a case where both instabilities are triggered\footnote{This process has been proposed to be used as a standard candle in \cite{Di:2024tlz}. It is also possible for axion cloud over-densities to collapse in the vicinity of rotating black holes, producing a similar explosive phenomena \cite{Yoshino:2012kn}. }. Lower intensity events can also take place due to excess axion accretion onto a near-critical axion star, which may be detectable at observatories such as SKA \cite{Maseizik:2024uln}.

Using numerical methods, in this work we explore the details of photon parametric resonance in the $\pi$-axiverse for generic physical processes coming from an unstable supercritical, over-density of $\pi$-axion dark matter\footnote{We note that some of the phenomenology discussed in \cite{Boskovic:2019qao} is very similar to the physics in this paper, albeit for a single axion model.}. Although we focus on the $\pi$-axiverse \cite{Alexander:2023wgk} as a benchmark model, our results are applicable to general multi-axion-dilaton dark matter models with high-dimensional parameter spaces. In the $\pi$-axiverse, there are in general four different interaction vertices between the $\pi$-axion and the photon, and resonance further depends on the $\pi$-axion masses $m_{\pi_i}$, the number of unique states, and dark matter over-density $\rho_{DM}$. In certain regions of the parameter space, we observe efficient photon production. Depending on the region, we find narrow \textit{or} broad resonance, spanning from radio up to infrared frequencies.

This papers is organized as follows. In Sec. \ref{sec:ResonanceEssentials}, we review the basics of parametric resonance, its relation to particle production in cosmology, and discuss the complexities of resonance with multiple oscillators. In Sec. \ref{sec:Model}, we provide the broad strokes of the $\pi$-axiverse while in Sec. \ref{sec:PhotonParReso}, we overview the photon interactions and set up the parametric resonance equations for the photon field amplitudes. Sec. \ref{sec:NumericalApproach} is devoted to describing the numerical approach to integrating the photon dynamics and the assumptions we make in the parameter space. Sections \ref{sec:Observability} and \ref{sec:Detectability} are the analysis of the detectability of the photon emissions, with a focus on MeerKAT, the SKA, and ngVLA. Here, we identify regions of the parameter space where the telescopes can resolve the transient signal associated with mergers and estimate limits on the population of the ($\pi$-)axion stars that can be derived by non-observation. We discuss our results and conclude in Sec. \ref{sec:Conclusion}. 


\section{Essentials of parametric resonance}\label{sec:ResonanceEssentials}

The phenomenon of parametric resonance is generic to many physical systems, from resonating a wine glass to amplifying LC circuits. Parametric resonance as a mechanism for exponential particle production has been the subject of a rich body of literature both for axion and dark matter physics \cite{Boskovic:2019qao, Dror:2018pdh,Brandenberger:2023idg,Amin:2023imi,Alonso-Alvarez:2019ssa,Lee:1999ae} and more broadly in cosmology and astrophysics \cite{Brandenberger:2022xbu,Fujisaki:1995dy,Fujisaki:1995ua}. We will here briefly cover the basics of parametric resonance in the simplest scenario of a single driven oscillator, before discussing the complexities of adding just a second driving force. This will serve to set the stage for the photon-$\pi$-axion interactions discussed in Sec. \ref{sec:PhotonParReso}.

For a scalar $\chi$ of mass $m_\chi$ experiencing an oscillating driving force, the equation of motion in momentum space reads
\begin{equation}
    \ddot{\chi}_k+(k^2+m_\chi^2+A\sin(\omega t))\chi_k=0, \label{eq:basic-eqmo}
\end{equation}
where $A$ and $\omega$ are constants with $\omega\neq m_\chi$. Such an equation can be understood in the field theory context as coming from a cubic vertex between two scalars, 
\begin{equation}
    \mathcal{L}=\frac{1}{2}\Big((\partial\chi)^2+(\partial\phi)^2-m_\chi^2\chi^2-m_\phi^2\phi^2-\lambda\phi\chi^2\Big),
\end{equation}
which would then correspond to $\omega=m_\phi$ and $\phi(t)=\phi_0\sin(m_\phi t)$ with $A=\lambda\phi_0$. The field $\chi$ experiences a periodically changing frequency $\omega(t)^2=k^2+m_\chi^2+\lambda\phi_0\sin(m_\phi t)$, which leads to resonance for momentum modes with only certain values of $k$. By changing variables to $m_\phi t=2z-\pi/2$, the equation of motion reduces to
\begin{equation}\label{Matt}
    \chi_k''+(A_k-2q\cos2z)\chi_k=0,
\end{equation}
where $A_k=4\frac{k^2+m_\chi^2}{m_\phi^2}$, $q=\frac{2\lambda\phi_0}{m_\phi^2}$, and the prime denote derivatives with respect to $z=m_\phi t/2$. Equation (\ref{Matt}) is the very well-known Mathieu equation \cite{MacLachlan}, which has been studied in many contexts. When a given momentum mode experiences resonance, its amplitude behaves as $\chi_k(t)\sim e^{\mu_k t}$, where $\mu_k$ is called the Floquet exponent. Whether a given mode experiences resonance or not depends on the relative values of $A_k$ and $q$. The study of the stability of such solutions is called Floquet theory, and the quasi-stable solutions are known as Floquet states \cite{Tsuji:2023dar}. For the Mathieu equation, a stability/instability chart nicely illustrates the behavior of the solutions in regions where amplification may or may not occur. An example of such a chart can be found in Fig. 1 in \cite{Allahverdi:2010xz}.

Studying the behavior of solutions to (\ref{Matt}) generally requires numerical approaches. The situation becomes radically more complex even in the case where there is a second driving force with different amplitude or frequencies, or if there are further nonlinear or friction interactions \cite{GHOSECHOUDHURY201485}, i.e. 
\begin{equation}\label{NonLinMatt}
    \begin{split}
        \ddot{\chi}_k+(A_k+q_1\sin(\omega_1 t)+q_2\sin(\omega_2 t))\chi_k&=0,\\
        \ddot{\chi}_k+c\dot{\chi}_k+(A_k+q\sin(\omega t))\chi_k&=0, \\
        \ddot{\chi}_k+(A_k+q\sin(\omega t))\chi_k+c\chi_k^3&=0.
    \end{split}
\end{equation}
An exhaustive study of such scenarios was carried out in \cite{Kovacic_2018}. In that work, the authors were able to generate analogs of the stability chart, Fig. 1 in \cite{Allahverdi:2010xz}, which required a significant amount of computing power. The resulting charts exhibit extremely nontrivial structure at the boundaries between stable and unstable regions (\textit{e.g.} Figs. 14 and 15 in \cite{Kovacic_2018}). As the analog of the parametric resonance equation for the multi-axion-dilaton model considered in this paper is significantly more complex \cite{Alexander:2023wgk} than any of (\ref{NonLinMatt}), we will result to exploring the resonance structure via a systematic exploration of the multidimensional parameter space of the model.

Once a numerical (or approximate analytical) solution is obtained for $\chi_k(t)$, it is straightforward to connect to particle production. The number density of the $\chi$ field is obtained by considering a Bogoliubov transformation in the WKB approximation \cite{Allahverdi:2010xz} or simply the total energy of the mode divided by the energy of each particle \cite{Kofman:1997yn},
\begin{equation}
    n_k(t)=\frac{\omega_k}{2}\Big(\frac{|\dot{\chi}_k|^2}{\omega_k^2}+|\chi_k|^2\Big)-\frac{1}{2}.
\end{equation}
The Bunch-Davies initial condition near $t=0$ is $\chi_k\approx\frac{1}{\sqrt{2\omega_k}}e^{i\omega_kt}, \  \dot{\chi}_k\approx i\sqrt{\frac{\omega_k}{2}}e^{i\omega_kt}$ guarantees $n_k(t\approx 0)=0$.  Clearly, for modes that grow as $|\chi_k(t)|\sim e^{\mu_kt}$, we have exponential production $n_k(t)\sim e^{2\mu_k t}$. The total occupation number is then the integral over all momenta.

The phenomena of parametric resonance is further relevant in many astronomical systems, and can potentially play a role in explaining anomalous electromagnetic signals across a wide range of frequencies. In particular, it has been demonstrated that a localized population of a single species of coherently oscillating ALP, such as a QCD axion, can spontaneously produce large numbers of photons at a resonant frequency when above a critical density threshold. For typical QCD axion masses, the resonant frequency falls in the radio band (MHz-GHz). This behavior is sensitive to the local dark matter (DM) energy density, and a denser configuration of axions such as an axion star is an ideal candidate for resonant photoproduction. Below a critical axion number density, the stimulated decay of axion dark matter into photons cannot sustain itself, and exponential resonance will not occur.  Nevertheless, axion stars and other non-critical but appreciably dense axion dark matter configurations are capable of injecting significant amounts of energy into propagating photons through stimulated decay alone. Further discussion on parametric resonance in axion stars can be found in \cite{Hertzberg:2018zte,Escudero:2023vgv,Zhang:2018slz,Visinelli:2017ooc,Amin:2020vja,Chung-Jukko:2023cow,Chung-Jukko:2024hod,Levkov:2020txo}, while \cite{BetancourtKamenetskaia:2024gcv} additionally considers potential connections to gamma ray bursts (GRBs). For further discussion on observable effects of non-resonant stimulated decay in axion stars, see \cite{Liu:2023nct,Qin:2023kkk,Ghosh:2020hgd,Arza:2019nta,Sun:2023gic}. Resonant photoproduction inside neutron stars (NS) has also been discussed as a production mechanism for observed NS flares \cite{Campos:2005eu}. Parametric resonance induced by gravitational waves has been studied in the context of ALP production near primordial black holes \cite{Boskovic:2019qao}, as a means of direct photoproduction \cite{Brandenberger:2022xbu}, and as a mechanism for rapid decay of ALP clumps \cite{Brandenberger:2023idg,Sun:2020gem}.

While the phenomenon of parametric resonance and systems of coupled oscillators have both been the subjects of rich fields of study on their own, there has been little investigation into the phenomenology of coupled parametric oscillators, and their resonance conditions in non-trivial systems. It has been shown in \cite{Copelli_2001} that in the event of identical oscillators, the system can be linearized and solved through similar means to traditional Floquet analysis, via construction of a matrix operator. In this case, it is found that the relative differences in phases of the oscillators can have significant effects on the resulting stability of the system, and so varying the relative phases of coupled parametric oscillators alone can result in a stable region of parameter space becoming unstable, or vice versa. This significance of relative phase difference on the stability of a solution across the entire range of possible parameters has been explored in \cite{McMillan_2022}, with the additional discovery that couplings between oscillators can also produce non-negligible corrections to the final regions of stability. Large systems of coupled parametric oscillators have also been studied in the context of condensed matter systems and generalizations of the Ising model \cite{Eichler_2023,Calvanese_Strinati_2019,Maddi_2022,Calvanese_Strinati_2020}.


Parametric resonance within a system containing multiple oscillators with similar order couplings and incommensurate relative frequencies is referred to as \textit{quasiperiodic} resonance, generically described by a differential equation of the form
\begin{equation}
    \ddot{\chi}_k + \Big(A_k + \sum_{i=1}^{N} q_{i}\sin(\omega_{i} t)\Big)\chi_k = 0.
\end{equation}
While for a single oscillator this reduces to a simple Mathieu equation, including additional oscillators lacks a closed-form solution. The specific case of $N = 2$ oscillators has been studied in \cite{Kovacic_2018} through direct numerical simulation, and  in \cite{Sharma_2017} where the authors attempted to employ a traditional Floquet analysis method by approximating the system as  periodic with a large principal period. The intractability of quasiperiodic systems within traditional analysis techniques has led to the development of alternative methods in pure mathematics, such as Kolmogorov-Arnold representation theory \cite{Tikhomirov1991OnTR, Schmidt_Hieber_2020}, which has even found application in machine learning \cite{liu2024kankolmogorovarnoldnetworks}.

One can additionally extend the case of constant or periodically driven systems to consider systems subjected to a quasiperiodic driving force, 
\begin{equation}
    \ddot{\chi}_k + \Big( \sum_{j=1}^{N} c_{j}(t)\Big)\dot{\chi_k} + \Big(A_k + \sum_{i=1}^{N} q_{i} \sin(\omega_{i} t)\Big)\chi_k = 0,
\end{equation}
which can be shown to introduce non-trivial -- and at times fractal -- modifications to the steady-state behavior of the system \cite{Verdeny_2016,Cadez_2017,JORBA1992111}. In many-body quantum systems, quasiperiodic driving forces can induce anomalous phases of matter in the system \cite{Ray_2019,Cubero_2012}, which can be described through a Renormalization Group approach, where fixed points in the theory correspond to exactly solvable models \cite{Gon_alves_2023,Feudel_1995}. The asymptotic behavior of quasi-periodically driven systems can be highly sensitive to variations in parameter values or initial conditions.

In the case of the $\pi$-axiverse, with a non-trivial number of fields, attempting to find an analytical approach will not be enlightening or practical. We will instead take a statistical approach to predicting resonance conditions from the results of numerical simulations. The $\pi$-axiverse, in its most general form, predicts equations of motion that behave as a series of coupled, damped/driven, quasiperiodic, parametric oscillators. Much analysis of axion-photon parametric resonance focuses on a single ALP\footnote{In a recent paper \cite{Zhang:2025cgt}, a multiple axion universe was considered within the context of structure formation.}. The rich phenomenology of parametric resonance within multi-axion models has not yet been thoroughly explored in the literature.

\section{$\pi$-axiverse from dark QCD}\label{sec:Model}
The benchmark model we will study in this work stems from a dark copy of the Standard Model (dSM) with gauge structure $SU(3)_C\times SU(2)_W\times U(1)_Y$, as in \cite{Alexander:2023wgk}. This differs slightly from the model in \cite{Alexander:2024nvi}, which assumes no weak sector. The primary distinction between the two models is the presence of flavor changing currents, which ultimately enriches the interacting structure of the relic DM candidates. The key feature of the model is an assumption of two energy scales; the UV scale setting both the dark Higgs vacuum expectation value (vev) $v$ and the QCD scale $\Lambda_{\text{dQCD}}$ taken to be 
\begin{equation}
    v,\Lambda_{\text{dQCD}}\gtrsim 10^{11} \ \text{GeV},
\end{equation}
which will additionally set the energy scale of the axion decay constant, and an IR scale corresponding to the dark quark masses, 
\begin{equation}
    m_q\ll \text{ eV}.
\end{equation}
The latter naturally follows from small Yukawa couplings. The dark sector communicates to the Standard Model (SM) from standard kinetic coupling \cite{Cline:2021itd} between the dark photon and SM photon\footnote{In \cite{Alexander:2023wgk}, we discuss in depth the production of the $\pi$-axions and portals to the visible sector. Models of this type may lead to the production of ultra-heavy states such as dark glueballs \cite{McKeen:2024trt,Athenodorou:2020ani}, which can be problematic. We assume that $F_\pi > H_{inf}, T_{max}$, where $H_{inf}$ is the Hubble scale at inflation and $T_{max}$ is the maximum temperature of the Standard Model plasma. This prevents freeze in/freeze out of the heavy dark QCD states. Such states may also simply be underproduced as described in \cite{Carenza:2022pjd}, or diluted due to entropy injection \cite{Halverson:2016nfq}. }.

The dark QCD sector with $N_f$ flavors is given by the Lagrangian 
\begin{equation}
\label{eq:dQCD}
    \mathcal{L}_{\mathrm{dQCD}}=-\frac{1}{2}{\rm Tr} \, G_{\mu\nu}G^{\mu\nu}+\sum_{i,j=1}^{N_f}\bar{q}^{i}(i\slashed{D}\delta_{ij}-m_{ij})q^{j},
\end{equation}
where $G_{\mu \nu}$ is the non-Abelian field strength tensor and the $q^i$ are Dirac fermions in the fundamental representation of the the dark $SU(N_c)$ gauge symmetry. The mass matrix $m_{ij}$ is in general an $N_f\times N_f$ matrix with off diagonal elements. The dark quarks are endowed with a millicharge under the SM $U(1)_{\text{EM}}$, which arises through kinetic mixing between the dark and SM photon \cite{Cline:2021itd}.

This theory exhibits a global $U(N_f)\times U(N_f)$ chiral symmetry, which is broken at low energies below the confinement scale $\Lambda_{\rm dQCD}$. In particular the breaking of the subgroup,
\begin{equation}
    SU(N_f)_{\text{L}}\times SU(N_f)_{\text{R}}\rightarrow SU(N_f)_V,
\end{equation}
leads to a spectrum of Goldstone bosons known as pions. In this work we consider the spectrum of pions as an effective `axiverse', analogous to the string theory axiverse \cite{Svrcek:2006yi,Arvanitaki:2009fg,Cicoli:2012sz}. At leading order, the $\pi$-axions may be described by a $\sigma$-model, (see \cite{Kaplan:2005es,Scherer:2005ri})
\begin{equation}\label{sigmamodel}
    \mathcal{L}_{\text{eff}}=\frac{F_\pi^2}{4}\text{Tr}[(D_\mu U)(D^\mu U)^\dagger]+\frac{\langle q\bar{q}\rangle}{2}\text{Tr}[MU+M^\dagger U^\dagger],
\end{equation}
where $F_\pi$ is the $\pi$-axion decay constant, $M$ is the mass matrix, $U={\rm exp}(\frac{2\pi i}{F_\pi}\pi^a T_a)$, with $T_a$ the generators for $SU(N_f)$, $\pi^a$ the $N_f^2-1$ dimensional  vector of $\pi$-axions, and $\langle q\bar{q}\rangle$ is the condensate parameter. 

While $F_{\pi}$, $\langle q \bar{q} \rangle$, and $\Lambda_{\rm d QCD}$, all arise from the physics of confinement, their precise relation is renormalization scheme dependent.  We will approximate the decay constant $F_{\pi}$ by the dQCD scale
\begin{equation}
    F_{\pi} \sim \Lambda_{\rm d QCD},
\end{equation}
though we note this relation is 
expected to hold only up to ${\cal O}(1)$ numerical factors, that can be precisely computed on the lattice. The $\pi$-axion masses, on the other hand, depend on both the quark mass scale and dark QCD scale, according to the Gell-Mann--Oakes--Renner (GMOR) relation \cite{Gell-Mann:1968hlm},
\begin{equation}\label{GMOR}
    m_\pi^2\simeq\frac{\langle q\bar{q}\rangle}{F_\pi^2}\sum_{i} m_{q_i}.
\end{equation}
where $m_{q_i}$ are the quark masses that constitute the $\pi$-axion in question and $\langle q \bar{q}\rangle \sim \Lambda_{\rm dQCD}^3$ is the quark condensate. The GMOR approximation (\ref{GMOR}) is less accurate when a given $\pi$-axion quark composition is the sum of more than two flavors, such as the SM $\eta$ particle, whose flavor eigenstate involves the up, down, and strange quarks. Even in this case, Eq.~\eqref{GMOR}  matches the observed mass up to a discrepancy of $4.6\%$, which is sufficient for our purposes. By assuming $\langle q \bar{q}\rangle \sim \Lambda_{\rm dQCD}^3\sim F_\pi^3$, we can approximate the $\pi$-axion masses by $m_\pi^2\simeq F_\pi\sum_im_{q_i}$.

As in \cite{Alexander:2023wgk,Alexander:2024nvi}, we will choose to  parameterize the quark masses with a characteristic mass for each \textit{generation} of up-type and down-type quark, $m_I,m_{II},...$, and use constants $c_1,c_2,...$ to distinguish quark masses in each generation. For example, for the $SU(6)$ case explored in \cite{Alexander:2023wgk}, we have six quarks in three generations (up, down, strange, charm, bottom, top), and write
\begin{equation}\label{SU6masses}
    \begin{split}
        m_u=c_1m_I, \ \ \ m_d=c_2m_I, \ \ \ m_s=c_3m_{II}, \ \ \ m_c=c_4m_{II}, \ \ \ m_b=c_5m_{III}, \ \ \ m_t=c_6m_{III},
    \end{split}
\end{equation}
assuming $m_I\leq m_{II}\leq m_{III}$ without loss of generality. The analog of the Standard Model $\pi_0$, which is composed of one up and one down quark, in our notation will have a mass $m_{\pi_0}\simeq\sqrt{F_\pi m_I(c_1+c_2)}$.

For the charged $\pi$-axions, the mass formula gets corrected by photon loops \cite{Kaplan:2005es}, both visible and dark. The charged $\pi$-axion mass can then be written as 
\begin{equation}
\label{eq:mpm}
    m_{\pi_i^{\pm}}^2 \simeq 
    \begin{cases}
    m_{\pi_i ^0}^2 + 2\xi_i {e'}^2 F_{\pi}^2 \,\,\, ,\,\, m_{\gamma'} < F_{\pi} \\
m_{\pi_i ^0}^2 + 2\xi_i e^2\varepsilon^2 F_{\pi}^2 \,\,\, ,\,\, m_{\gamma'}> F_{\pi} ,
\end{cases}
\end{equation}
where the two cases correspond to the dark photon dominating the loop correction vs. the dark photon being integrated out, and the loop being dominated by a visible sector photon. Here $e = \sqrt{4 \pi \alpha_e}$ is the dimensionless electric charge parameter, $e'=\sqrt{4\pi\alpha_{e'}}$ is the dimensionless dark EM coupling (where $\alpha_{e'}$ need not be $\sim\frac{1}{137}$), and $\xi_i$ is an $O(1)$ parameter unique to each $\pi$-axion, which can be positive or negative. Since the energy scale of $F_\pi$ is so large, in the case where $\varepsilon\sim 1$, the charged $\pi$-axions will be superheavy and subsequently will have decayed in the early universe. The unit millicharge scenario, therefore is such that the late-time spectrum of dark matter states are neutral, avoiding bounds on charged dark matter models \cite{Alexander:2023wgk}.
\section{Photon parametric resonance}\label{sec:PhotonParReso}

Experimental searches for axions are predicated on interactions with the Standard Model photon and fermions. The $\pi$-axiverse is characterized by a mixture of parity-odd and parity-even couplings, as one might conventionally associate with an axion field and a dilaton field, respectively. We consider four different vertices such that $\mathcal{L}_{\rm int}=\sum_{i=1}^4\mathcal{L}_{\text{int}}^{(i)},$ where, momentarily omitting the sum over species,
\begin{eqnarray}\label{interaction1}
\mathcal{L}_{\text{int}}^{(1)}&&=\frac{\lambda_1}{2 F_{\pi}}\varepsilon^2(\pi^0)F_{\mu\nu}\tilde{F}^{\mu\nu},\\
\label{interaction2}
\mathcal{L}_{\text{int}}^{(2)}&&=\frac{\lambda_2}{2} \varepsilon^2 (\pi^{+})(\pi^{-}) A_\mu A^\mu, \\
\label{interaction3}
\mathcal{L}_{\text{int}}^{(3)}&&=\frac{\lambda_3}{2 \Lambda_3 ^2} \varepsilon^2 (\pi^{+})(\pi^{-}) F_{\mu\nu}F^{\mu\nu}, \\
\label{interaction4}
\mathcal{L}_{\text{int}}^{(4)}&&=\frac{\lambda_4}{2 \Lambda_4 ^2} \varepsilon^2(\pi_i)(\pi_j) F_{\mu\nu}F^{\mu\nu}. 
\end{eqnarray}
Here, ${\cal L}^{(1)}$ is the standard pion-photon coupling resulting from the triangle anomaly, and ${\cal L}^{(2)}$ is the gauge covariant derivative of scalar QED, corresponding to the millicharges of the charged $\pi$-axions. Meanwhile ${\cal L}^{(3)}$ and ${\cal L}^{(4)}$ arise  as effective field operators arising from integrating out the heavy degrees of freedom in the dark Standard Model, the former interaction describing all charged $\pi$-axions and the latter describing all neutral (real and complex) $\pi$-axions. As in the visible SM, the weak bosons are electrically charged and can couple to the photon as a result. Therefore at high energies, two neutral $\pi$-axions can interact with photons through a loop process involving weak bosons as in $K-\bar{K}$ mixing in the visible SM. The photons could be emitted by the dark $W^{\pm}$ or possibly by the off-shell quarks.  A diagram illustrating this process is shown in Fig. 1 in \cite{Alexander:2023wgk}. From this we can estimate the energy scales $\Lambda_{3,4}$ as simply ${\Lambda_{3,4} } \sim m_{W} = g v$. The couplings $\mathcal{L}^{(3)}$ and $\mathcal{L}^{(4)}$ share similarities with couplings between the string theory dilaton $\Phi$ and the gauge field kinetic energy, which is generically of the form $\sim f(\Phi) F^2$ (\textit{e.g} \cite{Adshead:2023qiw,Cho:1975sf,Cho:1975sw}). In difference from the standard string theory dilaton, our effective vertices are four point, involving two $\pi$-axions and the photon kinetic energy. Nonetheless, the $\pi$-axiverse is very reminiscent of a multi-axion-dilaton model.

Finally, we consider the impact of the coherent oscillations of the $\pi$-axion fields on excitations of the Standard Model photon field. As discussed in \cite{Marsh:2015xka}, the axion couplings to photon can lead to a parametric resonance of the latter, which is enhanced in dense environments. This phenomenon is well studied in the case of conventional axions, see Refs.~\cite{Visinelli:2017ooc,Amin:2020vja,Amin:2021tnq,Du:2023jxh,Escudero:2023vgv,Chung-Jukko:2023cow}, where it can allow for indirect detection via various astrophysical signatures.
Parametric resonance has also been studied in the context of ultralight millicharged dark matter in \cite{Jaeckel:2021xyo}. Additionally, the gravitationally-induced resonance of axion fields and gravitons has been studied in  \cite{Alsarraj:2021yve,Brandenberger:2023idg}.  Here we study the possibility of parametric resonant production of photons in the $\pi$-axiverse.

Both parity-odd and parity-even portals will contribute to parametric resonance of photons.  Including the sum over $\pi$-axion species, the Lagrangian including all interactions for the SM photon is 
\begin{equation}\label{LU(1)}
\begin{split} 
   & \mathcal{L}^{\mathrm{int}}_{U(1)_\text{EM}}= \\
   &-\varepsilon^2\Big(\frac{\lambda_3}{2\Lambda_3^2}\sum_{i,j}^{\text{charged}}\pi_i \pi_j^*+\frac{\lambda_4}{2\Lambda_4^2}\sum_{i,j}^{\text{neutral}}\pi_i\pi_j^*\Big)F_{\mu\nu}F^{\mu\nu} \\
    & -\frac{\lambda_1}{2F_\pi}\varepsilon^2\sum_i^{\text{neutral,}\mathbb{R}}\pi_iF_{\mu\nu}\tilde{F}^{\mu\nu}-\frac{\lambda_2\varepsilon^2e^2}{2}\sum_{i,j}^{\text{charged}}\pi_i\pi_j^*A_\mu A^\mu \\
    &+\text{h.c.}
    \end{split}
\end{equation}
In the $\pi$-axiverse, there are states that are real (electrically neutral) scalars, and complex scalars that may be electrically charged \textit{or} neutral, the latter analogous to the Standard Model kaon. As was shown in \cite{Alexander:2024nvi}, for an arbitrary number of dark quark flavors $N_f$, the number of unique charged $N_\pm$, complex neutral $N_0^c$, and real neutral $N_0^r$ species when $N_f$ is even is
\begin{equation}\label{evenNstates}
  \text{even:} \ \ \  N_\pm=\frac{N_f^2}{4}, \ \ \ \ N_0^c=\frac{N_f(N_f-2)}{4}, \ \ \ \ N_0^r=N_f-1,
\end{equation}
while for $N_f$ odd,
\begin{equation}\label{oddNstates}
    \text{odd:} \ \ \ N_{\pm}=\frac{N_f^2-1}{4}, \ \ \ \ N_0^c=\frac{(N_f-1)^2}{4},\ \ \ \ N_0^r=N_f-1.
\end{equation}
The $\pi$-axions that experience the standard axion coupling to $F\tilde{F}$ are the $N_0^r$ real, neutral states. The total number of states includes the conjugates of the complex states, and it is trivial to check that they add up to the number of generators for $SU(N_f)$ as required, $2N_\pm+2N_0^c+N_0^r=N_f^2-1$ for $N_f$ even or odd. 

We take the Fourier representation of of the photon vector potential $\vec{A}$ as
\begin{equation}
    \vec{A}(t,\vec{x})=\sum_{\pm}\int\frac{d^3k}{(2\pi)^3}e^{i\vec{k}\cdot\vec{x}}\hat{\varepsilon}_{\vec{k},\pm}A_{\vec{k}\pm}(t)+\text{c.c.},
\end{equation}
where $\hat{\varepsilon}_\pm$ are the polarization vectors written in the circular basis satisfying $\vec{k}\cdot\hat{\varepsilon}_\pm=0$. For the $\pi$-axions, we assume the real and complex fields are in coherent states, with respective profiles \cite{Alexander:2023wgk}
\begin{equation}
    \begin{split}
        \pi_i^r(t)&=\pi_{i,0}^r\cos(m_it +\delta_i), \\
        \pi_j^c(t)&=\pi_{j,0}^ce^{i\theta_j}\cos(m_jt+\delta_j).
    \end{split}
\end{equation}
Here, $\pi_{i,0}^{r,c},\ \theta_i,$ and $\delta_i$ are all taken to be real numbers.
The equation for the photon mode functions is then of the form
\begin{equation}\label{modeEOM}
\begin{split} 
0&=    \Big(1+P(t)\Big)\Big(A''_{\pm}+k^2A_{\pm}\Big)+B(t)A'_{\pm} \\
&+\Big(C_\pm(t)k+D(t)\Big)A_{\pm},
\end{split} 
\end{equation}
where explicitly, 
\begin{equation}\label{Pcoeff}
    \begin{split}
         P(t)&=\frac{4\lambda_3}{\Lambda_3^2}\varepsilon^2\sum_{i,j}^{N_\pm}\pi^c_{i,0}\pi^c_{j,0}\cos(\theta_i-\theta_j) \cos\varphi_i(t)\cos\varphi_j(t)\\
        &+\frac{2\lambda_4}{\Lambda_4^2}\varepsilon^2\Big[2\sum_{i,j}^{N_0^c}\pi^c_{i,0}\pi^c_{j,0}\cos(\theta_i-\theta_j) +\sum_{i,j}^{N_0^r}\pi^r_{i,0}\pi^r_{j,0}\\
        &+4\sum_{i=1}^{N_0^r}\sum_{j=1}^{N_0^c}\pi_{i,0}^r\pi_{j,0}^c\cos\theta_j\Big] \cos\varphi_i(t)\cos\varphi_j(t)
    \end{split}
\end{equation}
\begin{equation}\label{Bcoeff}
    \begin{split}
        B(t)&=-\frac{4\lambda_3}{\Lambda_3^2}\varepsilon^2\sum_{i,j}^{N_\pm}\pi^c_{i,0}\pi^c_{j,0}\cos(\theta_i-\theta_j) \\
        &\times \Big(m_i\sin\varphi_i(t)\cos\varphi_j(t)+m_j\cos\varphi_i(t)\sin\varphi_j(t)\Big) \\
        &-\frac{2\lambda_4}{\Lambda_4^2}\varepsilon^2\Big[2\sum_{i,j}^{N_0^c}\pi^c_{i,0}\pi^c_{j,0}\cos(\theta_i-\theta_j) +\sum_{i,j}^{N_0^r}\pi^r_{i,0}\pi^r_{j,0}\\
        &+4\sum_{i=1}^{N_0^r}\sum_{j=1}^{N_0^c}\pi_{i,0}^r\pi_{j,0}^c\cos\theta_j\Big]  \\
        &\times \Big(m_i\sin\varphi_i(t)\cos\varphi_j(t)+m_j\cos\varphi_i(t)\sin\varphi_j(t)\Big),
        \end{split}  
\end{equation} 
along with
\begin{equation}\label{Ccoeff}
        C_{\pm}(t)=\pm\frac{2\lambda_1}{F_\pi}\varepsilon^2\sum_i^{N_0^r}\pi^r_{i,0}m_i\sin\varphi_i(t),
\end{equation}
defining the shorthand $\varphi_i(t)=m_it+\delta_i$. In both (\ref{Pcoeff}) and (\ref{Bcoeff}), the sum over the first line is the nine charged species, the other three sums are over the six complex and five real neutral species, while the sum in (\ref{Ccoeff}) is over the five real, neutral species. The $D(t)$ function is the sum over charged species,
\begin{equation}\label{Dcoef}
    D(t)=2\lambda_2\varepsilon^2e^2\sum_{i,j}^{N_\pm}\pi^c_{i,0}\pi^c_{j,0}\cos(\theta_i-\theta_j)\cos\varphi_i(t)\cos\varphi_j(t).
\end{equation}

 The differential equation (\ref{modeEOM}) with non-vanishing $P(t),B(t),C_\pm(t)$ and $D(t)$ is  significantly more complicated than the Mathieu equation generally encountered in axion cosmology. In the most general case where all of the $\pi$-axion masses differ, there will be an exceedingly rich resonant structure due to the incommensurate differences found in both the driving frequencies and the frequencies of the coherently oscillating amplitudes. However, even in the case where the dark quark masses are fully degenerate, the resultant $\pi$-axion masses will still produce a spectrum with incommensurate $\mathcal{O}(1)$ separation, and therefore the phenomenology of parametric resonance will always differ from the standard (single-field) axion scenario \cite{Arza:2020eik,Sigl:2019pmj,Tkachev:1987cd,Tkachev:2014dpa,Arza:2018dcy,Hertzberg:2018zte}. In scenarios where the frequency of the produced photons is observable, as in \cite{Amin:2020vja}, the $\pi$-axiverse predicts a distinguishable signal due to this resonance structure. The combined effect of these resonances can
also accelerate the instability of axion stars. 

The nature of the photon parametric resonance will naturally be sensitive to the density of dark matter in the environment where the interaction takes places. The dark matter density itself, $\rho_{DM},$ can be composed of an admixture of the $\pi$-axions in our model, i.e. $\rho_{DM}=\sum_i\rho^i_{DM}$. We will assume that the amplitudes of the $\pi$-axions are related to the individual species' mass and contribution to the dark matter density as \cite{VanTilburg:2015oza,Alexander:2023wgk}
\begin{eqnarray}
    \pi_{i,0}\simeq \frac{\sqrt{2\rho^i_{DM}}}{m_i}.
\end{eqnarray}
This assumption will allow us to explore the multidimensional parameter space of the $\pi$-axiverse, which includes the full set of parameters\footnote{Recall from the GMOR relation (\ref{GMOR}) that the $\pi$-axion mass relates to the dark quark masses as $m_{\pi_i}^2\sim \frac{\langle q\bar{q}\rangle}{F_\pi^2}m_{q_i}$. As in \cite{Alexander:2023wgk}, we take $\langle q\bar{q}\rangle\sim \Lambda_{dQCD}^3\sim F_\pi^3$, so that $m_{\pi}^2\sim F_\pi m_{q_i}$. Therefore the parameter space of the dark quark mass is easily related to that of the $\pi$-axion, up to the caveat of the values of the constants $c_i$.}
\begin{equation}
    \{\varepsilon,F_\pi,\Lambda_3,\Lambda_4,N_f,m_{q_i},\rho_{DM},c_i\},
\end{equation}
and in principle the $\delta_i$ and $\theta_i$. Only in certain regions of the parameter space will we find that parametric resonance occurs for the photon due to the oscillations of the $\pi$-axions.

Due to the size of the parameter space, we naturally will have to make some initial assumptions for our numerical analysis. We will set $\varepsilon=1$ so that the interactions are as significant as possible. As a result, the charged $\pi$-axions are superheavy, will have decayed in the early universe, and will not be a part of the relic dark matter. This greatly simplifies our analysis, since $\mathcal{L}_{int}^{(2)}$ and $\mathcal{L}_{int}^{(3)}$ no longer contribute to the photon interactions. We will additionally focus on the case of $N_f=3$, which will give us one complex neutral state, and two real neutral states. We therefore will not probe the parameter space of $\varepsilon$ or $\Lambda_3$, but systematically explore the parameter subspace of $\{F_\pi, \Lambda_4,m_{q_i},\rho_{DM}\}$. We  elaborate on the additional choices we make for sampling certain model parameters in the following section.

\section{Numerical approach and parameter space}\label{sec:NumericalApproach}
Here we present an outline of our numerical approach and an overview of the regions of parameter space explored. The dark quark masses are chosen in order to roughly mimic the structure of the quark mass spectrum in the Standard Model. Explicitly, they are defined as in (\ref{SU6masses}). The scaling constants $c_i$ are sampled from a uniform distribution over the interval [0.7, 1.3], in order to require an $\sim\mathcal{O}(1)$ separation between quark masses of the same generation. This is further described in Table \ref{tab:ci_table}. We restrict to the $SU(3)$ case, where the free parameters are $m_I$, $m_{II}$, and $c_{1-3}$. We further assume that $m_{II}=2m_{I}$, which is a tighter spectrum than the Standard Model quarks.

The ultra-heavy charged $\pi$-axion species do not have cosmologically stable lifetimes, therefore we assume their local energy density is negligible. Appreciable charged $\pi$-axion populations could still be produced in the late-universe, namely in the magnetospheres of pulsars \cite{Hook:2018iia,Noordhuis:2023wid}, which are also objects of interest concerning the production of FRBs \cite{Kyriazis:2022gvw,Pan:2022gwf}. Such interactions may still hold relevance within the $\varepsilon=1$ limit, but we will leave this analysis to future work. 
Note that a strict ordering in the hierarchy of the dark quark mass species is not imposed beyond the mass generation constants, $m_{I}<m_{II}$.




\begin{table}[h]
    \centering
    \begin{tabular}{|cl|c|c|}
    \hline
         $c_1$ & $\in \mathcal{U}[0.7, 1.3]$ & $u$ & $m_{I}$\\
         $c_2$ & $\in \mathcal{U}[0.7, 1.3]$ & $d$ & $m_{I}$\\
         $c_3$ & $\in \mathcal{U}[0.7, 1.3]$ & $s$ & $m_{II}=2m_I$\\
         $c_4$ & $= 0$ & $c$ & $m_{II}$ \\
         $c_5$ & $= 0$ & $t$ & $m_{III}$ \\
         $c_6$ & $= 0$ & $b$ & $m_{III}$ \\
    \hline
    \end{tabular}
    \caption{Sampling distributions of the theory constants $c_i$ for the $SU(3)$ like case, and their corresponding dQCD quark species and mass generation. For computational simplicity, we omit all bound states composed of the $c$, $t$, and $b$ dark quarks.}
    \label{tab:ci_table}
\end{table}

The local dark matter energy density is taken to be composed of an even split between real neutral and complex neutral populations, $
    \Sigma^{N_r}_i{(\rho_r)_{_i}} = \Sigma^{N_c}_i{(\rho_c)_{_i}}$,
where
\begin{equation}
    \rho_{_\text{DM}} = \sum^{N_r}_i{(\rho_r)_{_i}} + \sum^{N_c}_i{(\rho_c)_{_i}}.
\end{equation}
For the $SU(3)$ case, $N_r=2$ and $N_c=1$. The photon number density is given by
\begin{equation}
    n_k(t)=\frac{k}{2}\Big(|A_k|^2+\frac{|\dot{A_k}|^2}{k^2}\Big)-\frac{1}{2},
\end{equation}
while the initial conditions are
\begin{align}
    A_k(0)&=\sqrt{\frac{1}{2k}}, \ \ \ \ \ \ \ 
    \dot{A}_k(0)=\sqrt{\frac{k}{2}},
\end{align}
which has the consequence that $n(0)=0$. In practice, this means that resonance in our analysis is indicative of an initial $\pi$-axion population configuration which is above some critical point in the local population parameter space and is capable of undergoing resonance spontaneously. In other words, the parametric resonance is the result of a perturbed supercritical axion star. In the late universe, such systems may arise via axion star mergers \cite{Chung-Jukko:2024hod,Amin:2020vja,Du:2023jxh,Hertzberg:2020dbk}, accretion of local axionic DM onto a near-critical axion star, or collapse of a dilute axion star into a dense axion star \cite{Visinelli:2017ooc,Escudero:2023vgv}. We discuss non-spontaneous decay in the conclusion, Sec. \ref{sec:Conclusion}.

We then numerically integrate the $\pi$-axion-photon equations of motion (EOM) (\ref{modeEOM}) for a given choice of parameters from this initial condition, using a heuristic hybridization of \texttt{RK45} and \texttt{BDF} techniques to solve the photon field amplitude ODE in time for each Fourier $k$-mode independently. The potentially stiff nature (depending on the region of parameter space being considered) of the ODE resulting from the different oscillatory terms in the general $\pi$-axiverse EOM motivates the choice to employ this heuristic in order to balance numerical precision requirements while maintaining computational efficiency. For robustness, we solve the EOM from $k = 1$ to $k = 200$, in units of $m_0=\sqrt{F_\pi(c_1+c_2)m_I}$, \textit{the lightest $\pi$-axion species}, with a resolution of $\Delta k = 0.1$. For each $k$-mode, we numerically integrate from $t = 0$ to $t = 100$ in units of $m_0$ with a resolution of $\Delta t = 0.1$, automatically terminating integration if excessive conditions for resonance ($\gg\mathcal{O}(10^{10})$ increase in $n_k$) have been reached, or if the integration time in geometric units exceeds the maximum stable length scale for a homogeneous axionic structure in the associated region of parameter space. This secondary temporal cutoff is both imposed by the physical limits of the assumed system, and is well-motivated because different solutions to the quasiperiodic Mathieu-like equations are capable of entering and exiting regions of instability at different times as the system evolves; the asymptotic nature of the system is not always well-defined or necessarily informative of the stability of the system within timescales of physical interest. Therefore, numerical integration is both necessary due to a lack of analytical solutions for equations of this form, and phenomenologically preferred over asymptotic methods, such as the traditional Floquet-like approach. We also note that, in a departure from the single-species axion case, it is in general necessary to use a resolution of $\Delta k < 1$ and $\Delta t < 1$ to solve the Mathieu-like equations, to ensure we still capture the quasi-periodic dynamics involving $\pi$-axion species with masses non-commensurate to the reference unit mass.

We sample $N = 7,560$ points in parameter space, arranged as a hypercube with uniform spacing over the parameter ranges $\Lambda_4 \in [0, 40] \log_{10}{\text{GeV}}$, $\rho_{DM} \in [0, 40] \log_{10}{\text{GeV/cm}^3}$, $g_{\pi\gamma\gamma} \in [-30, -5] \log_{10}{\text{GeV}}$, and $m_{\pi} \in [-8, -4] \log_{10}{\text{eV}}$. Motivated by astrophysical surveys of axion-photon interactions, we elect to sample over the observable parameters $m_{\pi}$ and $g_{\pi\gamma\gamma}$ rather than the fundamental theory parameters $\left\{m_{q_i}, c_i, F_{\pi}\right\}$. The coupling coefficient between real neutral 
species, 
\begin{equation} 
g_{\pi\g\g} = \alpha_{\text{EM}}\frac{2 \lambda_1}{F_{\pi}}\varepsilon^2,
\end{equation}
is the $\pi$-axiverse generalized counterpart to the $g_{a\gamma\gamma}$ coupling constant governing single-field axion-photon interactions. Note that in the $\pi$-axiverse, the parity-even coupling to the $F^2$ term can induce oscillations in the SM fine-structure constant \cite{Alexander:2023wgk}. As an approximation we omit these corrections in our numerical investigation as they are typically negligible, but it should be noted that for regions of parameter space where $\Lambda_4$ is small ($\lesssim 10^8 \text{ GeV}$), some samples could be excluded due to additional constraints imposed from precision measurements of $\alpha_\text{EM}$. At each point in this parameter space, we invert the GMOR relations given in Equation (\ref{GMOR}) assuming median values of $c_i=1$, and the above definition of $g_{\pi\gamma\gamma}$ to obtain the corresponding dark quark masses $m_{q_i}$ and decay constant $F_{\pi}$. 

\begin{figure}[h]
    \centering
    \includegraphics[width=1.0\textwidth]{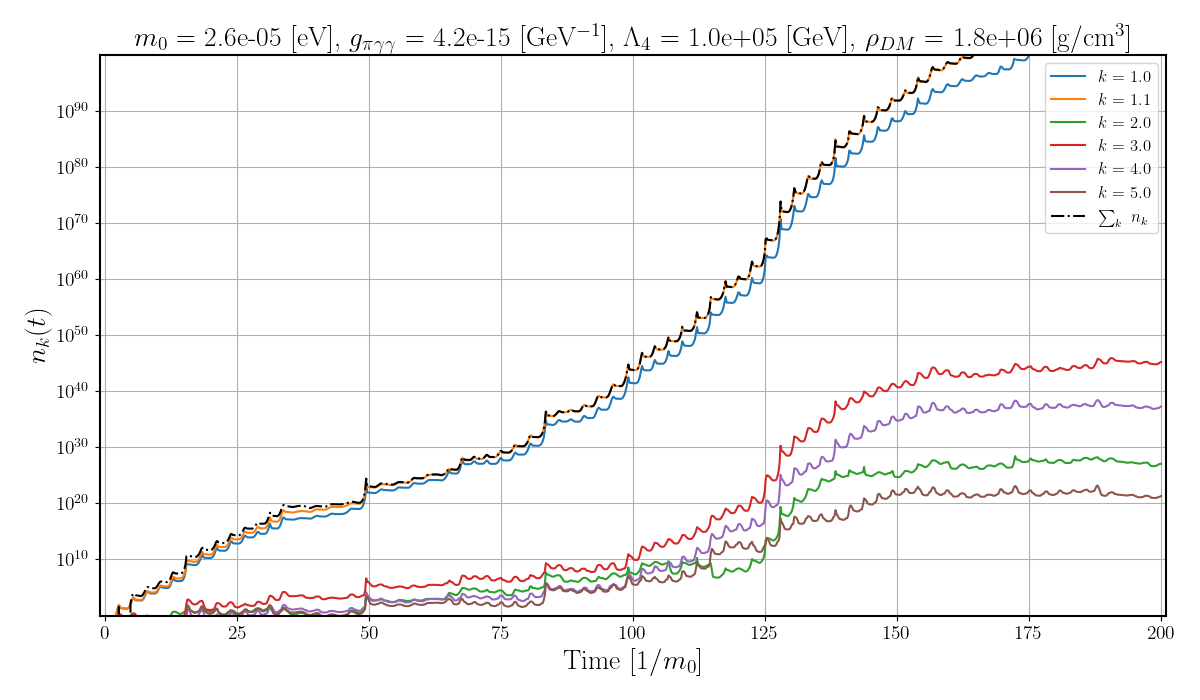}
    \caption{An illustrative example plot of the photon field occupation number as a function of time for a  narrow-band resonance event, showing both the evolution of a small sample of resonant $k$ modes, as well as for the total occupation number across all $k$-modes. 
    Note that here the peak frequency, $k\sim1.1 \,m_0$ is not an integer multiple of any of the fundamental $\pi$-axion masses (here, they are $\sim1 \,m_0$, $\sim1.2 \,m_0$, and $\sim1 .4\,m_0$). This particular example lies in the DALI experiment's projected sensitivity region of the $m_{\pi}$-$g_{\pi\g\g}$ plane \cite{DeMiguel:2024cwb}. Additionally, one can observe that the ordering of relative power between $k$-modes can change between early and late-time behavior.  
    The total time plotted covers only $5.025\times10^{-9}~\text{s}$, before factoring in any additional cutoffs motivated by physical length scale constraints or local DM energy density depletion. The peak frequency of the final spectrum is in the microwave band, $\lambda \sim 1.5 \text{ cm}$ corresponding to a frequency $\nu\sim 20$ GHz.}
    \label{fig:piaxi-nk-ex2}
\end{figure}

\begin{figure}[h]
    \centering
    \includegraphics[width=0.9\textwidth]{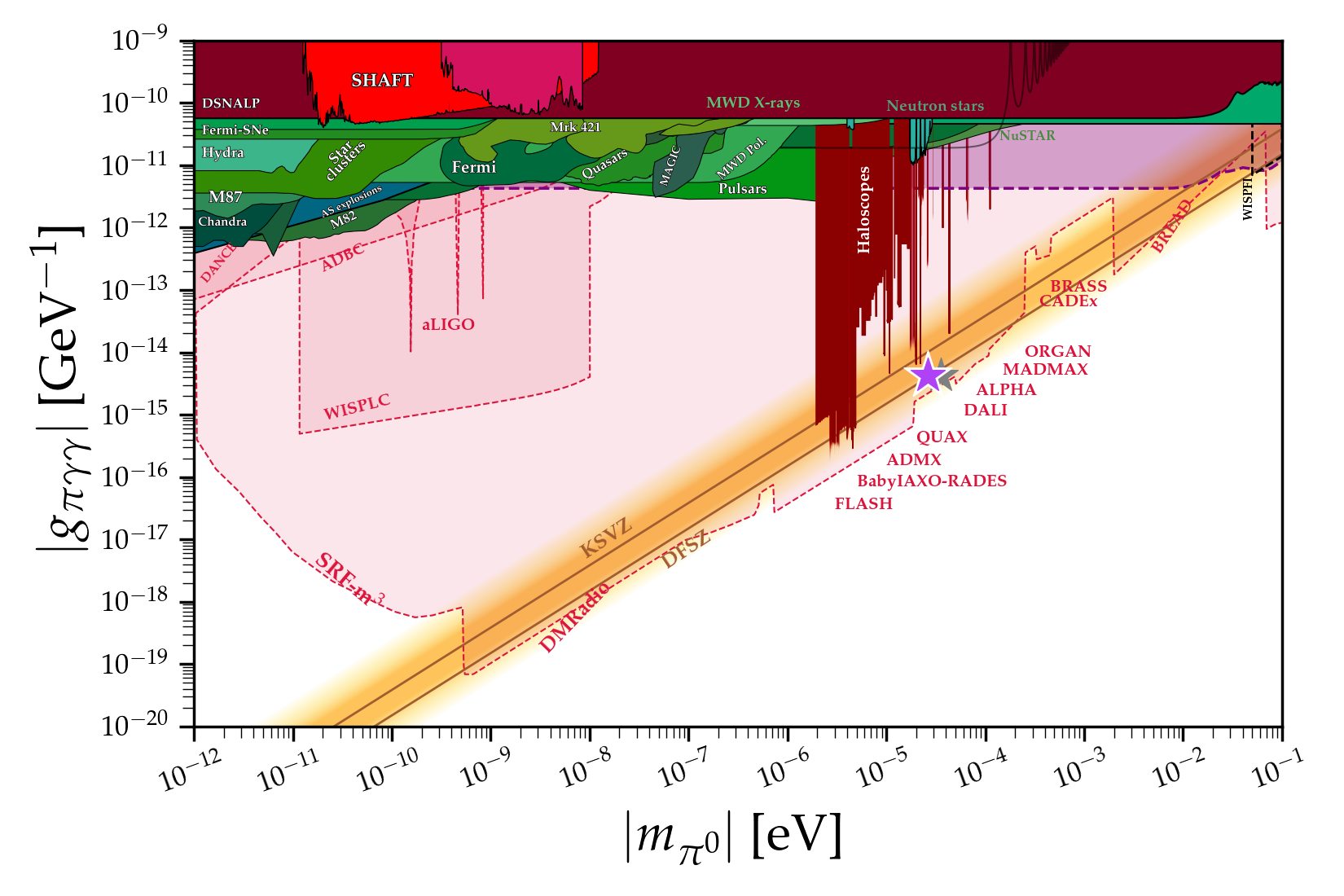}
    \caption{The $m_{\pi}$-$g_{\pi\g\g}$ search plane, with plotted constraints from single-species axion surveys. The purple star denotes the lightest real neutral $\pi$-axion species (which also defines the unit mass:  $m_0 \equiv\min[m_{\pi^0_i}]$), and the gray star is for the second, heavier, real neutral $\pi$-axion. Note that only the real neutral species participate in the interaction mediated by this coupling. This plotted example corresponds to the resonance spectrum plotted in Figure \ref{fig:piaxi-nk-ex2}. For further details on the experimental limits plotted here, refer to the AxionLimits software library \cite{AxionLimits}.}
    \label{fig:piaxi-alp-ex2}
\end{figure}

\begin{figure}[h]
    \centering
    \includegraphics[width=1.0\textwidth]{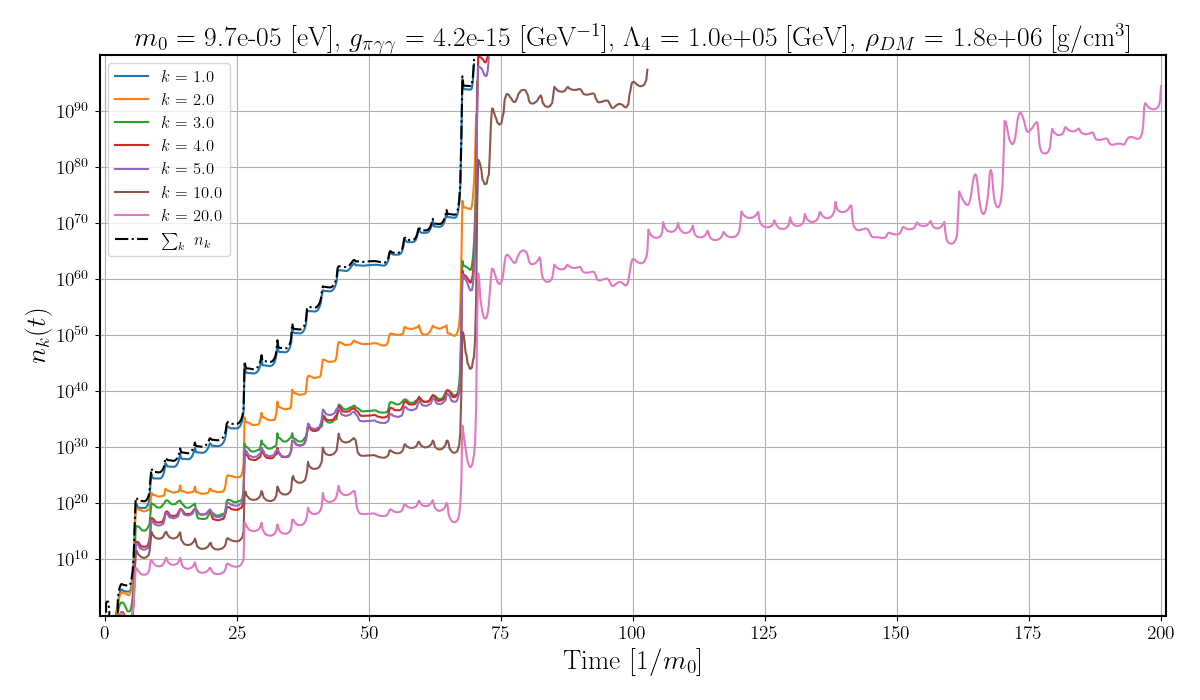}
    \caption{An example of a broad-band resonance spectrum 
    corresponding to $m_{\pi}^{(0)}=m_0=9.70\times10^{-5} \text{eV}$, $\Lambda_{4}=10^{5} ~\text{GeV}$ and $g_{\pi\gamma\gamma}=4\times10^{-15} ~\text{GeV}^{-1}$. In this case, the total time plotted covers only $1.36\times10^{-9}~\text{s}$, before factoring in any additional cutoffs motivated by physical length scale constraints or local DM energy density depletion. Note that while all $k$-modes exhibit resonant growth, the smaller modes grow more rapidly than larger ones, so in this case the spectrum is still  peaked at about $k=1$ ($\lambda \sim 1.4~\text{cm}$ or $\nu\sim 21$ GHz). Additionally, note that while all modes appear asymptotically unstable, the rate of photon number growth can vary greatly over time, with some sections demonstrating sharper instabilities than others.}
    \label{fig:piaxi-nk-ex3}
\end{figure}

For each point in parameter space, we repeat the integration three times with a different realization of uniformly sampled initial phase values, resulting in a total of $N = 22,680$ realizations of photon production spontaneously sourced by local $\pi$-axiverse populations. We additionally resample the constants $c_i$, leading to a tight clustering of final $m_{\pi}$ values about the hypercube points defined initially. We classify resonance at a given timestep $t$ if $n_k(t)/n_k(0) \gtrsim \mathcal{O}(10^{10})$, but in contrast to the procedure normally employed for single-species axion cases, we continue to integrate until we hit our timescale cutoff (or approach numerical precision limits). This is done in order to capture the phenomenology of the multi-species interactions, as they allow for the photon field to exit resonance after some finite amount of time passes, which could then result in the occupation number dropping below the resonance classification limit once again. We find that these events would result in finite but appreciable photon production, leading to a spectrum which is both lower in power and has a broader energy injection spectrum than typical transient events sourced by resonances. 

\section{Observability}\label{sec:Observability}

In principle the $\pi$-axion mass can span many decades in magnitude, meaning that photons produced by resonance range similarly in energy. However, very low energy photons will be absorbed by cosmic/atmospheric plasma. In addition, we aim to leverage the power of up-coming radio telescopes, like the SKA and ngVLA, in our study. To account for both the preceding points we focus on the mass range corresponding to signals from $50$ MHz to $50$ GHz. For single axion models this would equate to $m_\pi$ between $\sim 5 \times 10^{-7}$ and $\sim 10^{-3}$ eV. With this in mind, our primary question is: where we would observe parametric resonance between $\pi$-axions and photons? To answer this question we will study the growth rate of photon population under parametric resonance. In particular we study 
\begin{equation}
    \mu = \frac{d \ln{n}}{d t} \; , 
\end{equation}
and $n$ is the photon occupation number. For simplicity we can study this in the homogeneous case and apply it to cases with spatially varying $\pi$-axion density by comparing it to the rate photons escape from the structure in question~\cite{Hertzberg:2018zte}
\begin{equation}
    \mu_\mathrm{esc} \approx \frac{c}{D} \; ,
\end{equation}
where $D$ is the length-scale of the structure. Thus, we determine that structures of a size $D$ will experience a resonant instability if $\mu = \mu_H - \mu_\mathrm{esc}$ exceeds $0$ ($\mu_H$ is the homogeneous growth rate). We then scan the parameter space of $\{g_{\pi \gamma\gamma}$, $m_\pi$ (or $m_0$), $\rho_\pi$, $\Lambda_4\}$ to determine the instability regions. From this, we aim to infer emissions produced by mergers of objects just below the stability threshold. Once we have established the parameter space regions that result in observable mergers, we can consider the implications of merger non-observation with upcoming experiments. 

\begin{figure}
    \centering
    \resizebox{0.8\hsize}{!}{\includegraphics{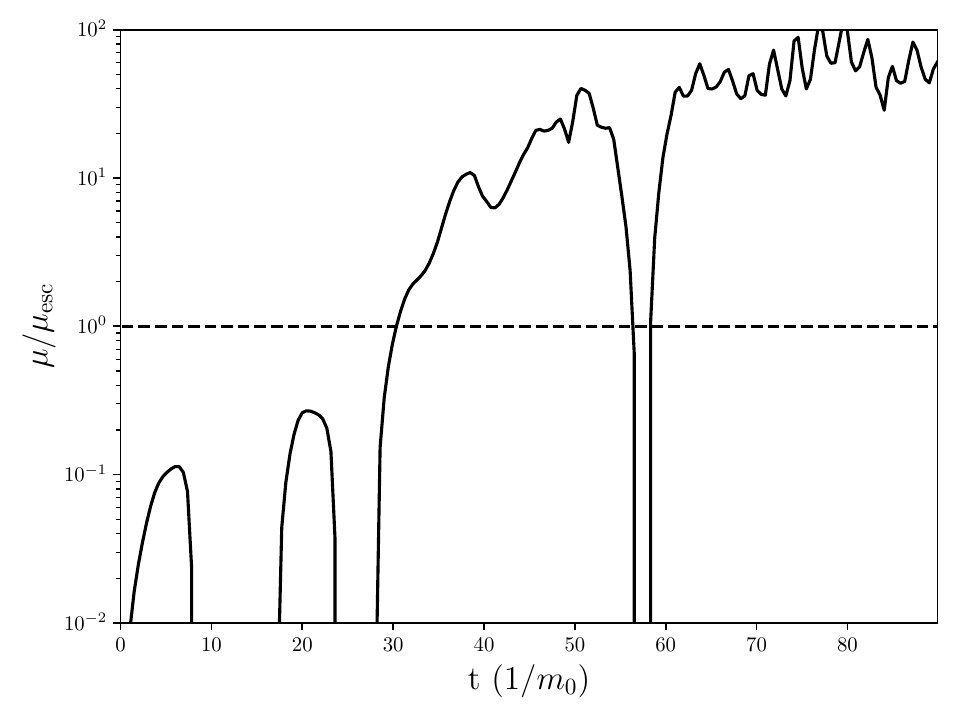}}
    \caption{The logarithmic growth rate for a simulation run where $g_{\pi\gamma\gamma} \sim 10^{-15}$ GeV$^{-1}$, $m^{(0)}_\pi=m_0 \sim 10^{-6}$ eV, and $\rho_\pi = 10^{25}$ GeV cm$^{-3}$.}
    \label{fig:growth-plot}
\end{figure}

We present an example homogeneous growth rate as a function of time, smoothed and normalized to $\mu_\mathrm{esc}$, for the case when $g_{\pi\gamma\gamma} \sim 10^{-15}$ GeV$^{-1}$, $m_\pi \sim 10^{-6}$ eV, and $\rho_\pi = 10^{25}$ GeV cm$^{-3}$ in Figure~\ref{fig:growth-plot}. From this plot we infer that structures at this density with $D \gtrsim 8 \times 10^{-2}$ km will experience resonant instability. It is interesting that a periodic cessation of growth appears to be a generic feature of $\pi$-axion models. For now, it is sufficient to note that the $\pi$-axion model experiences resonance but does not satisfy the single species requirement for resonance $g_{a\gamma\gamma} F_a > 0.3$~\cite{Hertzberg:2018zte}. This demonstrates the richness of a multi-axion phenomenology with both axion and dilaton-like couplings, as the aforementioned requirement restricts one to less than conventional axion theories in single-species scenarios. 

\subsection{Axion star mergers}

Using Fig.~\ref{fig:growth-plot} we can set a maximum stable size for $\pi$-axion structures, for a given set of $\{g_{\pi \gamma\gamma}$, $m_\pi$, $\rho_\pi$, $\Lambda_4\}$. Since we use a homogeneous density calculation, we will translate to the case of spatial variation by requiring that the average densities are equivalent within the structure, or
\begin{equation}
    \frac{3 \int^R_0 \rho_* (r) r^2 dr}{R^3} = \rho_{\pi} \; , 
\end{equation}
where $\rho_{\pi}$ is the homogeneous density. Thus, the knowledge of the stability threshold $R_\mathrm{max}$, in conjunction with an assumption of $\rho_* (r)$, allows us to determine the maximum mass for $\pi$-axion structures with a given average density. The presence of a maximum size for stability indicates that any increase beyond this mass, while not lowering the density, will result in the emission of radiation. Thus, we expect a build up of axion stars at such a threshold, with the only way to temporarily cross it being merger.  

Two types of merger need to be considered, delineated by whether the remnant has long-term stability or not. These states are separated by a density criterion, i.e. $\rho_\pi \gtrsim m_\pi^2 F^2$ indicates a compact star that will decay completely. Long-term stable remnants have a mass $\approx 0.7$ of the initial total~\cite{Hertzberg:2020dbk,Schwabe:2016rze} but will still radiate energy before relaxing to their stable state. For unstable remnants, around $0.4$--$0.5$ of the total pre-merger mass is radiated as photons~\cite{Chung-Jukko:2024hod}, about twice the stable remnant case~\cite{Hertzberg:2020dbk}. Thus, we expect all mergers to result in a sudden flash of radio frequency photons. The previously cited results~\cite{Hertzberg:2020dbk,Chung-Jukko:2024hod} apply to a single-field case, but should constitute reasonable estimates for multiple axions. However, it is worth mentioning cases where this estimate potentially fails. The first is when inter-species interactions differ greatly from self-interactions. Particularly, a total lack of inter-species interactions would potentially increase the density of axion stars and thus might alter the stability window. Second, there is the possibility of specific local inter-species phase configurations affecting the merger process. In our case, the inter-species and self interactions are of similar strength and we marginalize our results over local phase configurations. However, computational limitations mean that we do not have a large number of sampled phase configurations. Thus, these might still have an impact on our results.

To study the case of two merging $\pi$-axion stars with parameters matching our example from Fig.~\ref{fig:growth-plot}, we will employ the sech ansatz so that
\begin{equation}
    \rho_* (r) \propto \frac{\mathrm{sech}^2{x}}{R^3} \; ,
\end{equation}
where $R$ is taken to be the stability threshold, $x = \frac{r}{R}$, and this will be normalized via the average density condition. In this case, we find a mass of $M_{*,\mathrm{max}} = 1.29 \times 10^{37}$ GeV and radiated energy:
\begin{equation}
    E_{*,\gamma} \approx 2\delta_E M_{*,\mathrm{max}} \; ,
\end{equation}
where $\delta_E$ determines the photon radiation fraction. For a stable remnant, 
this yields a value of $E_{*,\gamma} \approx 5 \times 10^{36}$ GeV. This energy is likely to be emitted on a relatively short time scale~\cite{Hertzberg:2020dbk}. If we assume it matches $\frac{2 R}{c}$, we will find an intrinsic power on the order of $10^{41}$ erg s$^{-1}$. Thus, even for substantially slower emissions, this merger produces a flash with luminosity similar to a faint FRB~\cite{Luo:2020wfx}. Therefore, we will determine detectability in a fashion similar to FRBs. It is important to note that we are limited to studying cases where the density does not change during merger, as we lack a means to estimate the density change for arbitrary parameters. This is a highly improbable scenario, but should at least provide an estimate of expected emissions. This is because the resonant condition can be expressed as one on the axion number~\cite{Hertzberg:2020dbk}, which is not affected by our density assumption. The merger rate will be affected by this. However, the effect is difficult to determine, as larger pre-merger densities result in fewer stars per halo, but, more massive stars have higher merger probability. 

\subsection{Flux from transient events}

A central issue in the detectability of a transient event is the apparent duration. A relatively bright event could produce faint fluxes if the photon arrival time is sufficiently dispersed by scattering during propagation. Thus, we determine the flux from $\pi$-axion star mergers as follows:
\begin{equation}
    S_\mathrm{merge} = \frac{E_{*,\gamma}}{4\pi d_L^2 \tau_\mathrm{obs} \Delta \nu} g(\nu) \; ,
\end{equation}
where $\tau_\mathrm{obs}$ is the apparent duration of the burst, $d_L$ is the luminosity distance, $g(\nu)$ is the spectral shape function, and $\Delta \nu$ is the emission bandwidth. The observed duration can be computed as follows~\cite{2024ChPhL..41k9801W}
\begin{equation}
    \frac{\tau_\mathrm{obs}}{1+z} \approx \sqrt{\tau_\mathrm{int}^2 + \tau_\mathrm{host}^2 + \frac{\tau_\mathrm{DM}^2}{(1+z)^2}} \; ,
\end{equation}
where $\tau_\mathrm{int}$ is the intrinsic duration, $\tau_\mathrm{host}$ is the scattering delay from the host environment, and $\tau_\mathrm{DM}$ is the scattering delay from line-of-sight free electrons or~\cite{2024ChPhL..41k9801W}
\begin{equation}
    \tau_\mathrm{DM} = 8.3 \ \mu\mathrm{s} \left( \frac{\mathrm{DM}}{\mathrm{pc\ cm}^{-3}} \frac{\Delta \nu_\mathrm{MHz}}{\nu^3_\mathrm{GHz}}\right) \; ,
\end{equation}
where DM is the dispersion measure (integrated electron number density along the line of sight).
To find the bandwidth we use
\begin{equation}
    \Delta \nu \approx \sqrt{\Delta\nu_\mathrm{ins}^2+\nu^2\frac{\sigma_v^2}{c^2}} \; ,
\end{equation}
where $\sigma_v$ is the velocity dispersion of the parent halo and $\Delta\nu_\mathrm{ins}$ is the bandwidth of the observing instrument.
For axion stars within 1 Gpc we can assume $\tau_\mathrm{obs} \approx \tau_\mathrm{DM}$ and safely use $\mathrm{DM} \approx 900$ pc cm$^{-3}$. We use such a large distance to ensure a reasonable merger rate, as these can be as low as $\sim 10^{-18}$ per galaxy per year~\cite{Hertzberg:2020dbk}. For our model, we will determine the duration as the time required to receive 90\% of the total flux within the given instrument's frequency range. 

For the spectral shape $g(\nu)$, our approach is to take occupation number data for each mode $n_\nu(t)$ and build a spectrum using 
\begin{equation}
    g(\nu) = \eta\times  \mathrm{max}\left(\dot{n}_\nu\right) \; , \label{eq:spec}
\end{equation}
where $\eta$ is a normalization constant, as an estimate of the photon number at the given frequency. The maximum is taken within the crossing time $\frac{2 R}{c}$ for the merged remnant (after initial scalar radiation).

\subsection{Merger rate}
Once we have determined observable parameter regions, we need to estimate the regularity of merger occurrence. We will compute the rate of axion star merger within a spherically symmetric dark matter halo from~\cite{Hertzberg:2020dbk}
\begin{equation}
    \Gamma_{**,\mathrm{halo}} = 4\pi \int \frac{r^2}{2} \left(\frac{\rho_\mathrm{halo} f_*}{M_*}\right)^2 \langle \sigma_*(v) v \rangle dr \; ,
\end{equation}
where $\rho_\mathrm{halo}$ is the dark matter density of the parent halo (assuming a NFW profile), $f_*$ is the fraction of dark matter in the form of $\pi$-axion stars, $\sigma_*$ is the collisional cross section, and $v$ is the relative velocity between the merging objects. Note that the actual axion star distribution will be modified by tidal disruption, a full accounting for this is beyond the current work~\cite{kavanagh_stellar_2021} where the NFW halo will serve as a first approximation.
We compute $\sigma_*$ via
\begin{equation}
    \sigma_* (v) = 4 \pi R_*^2 \left(1 + \frac{v_{*,\mathrm{esc}}^2}{v^2}\right) \; ,
\end{equation}
where $v_{*,\mathrm{esc}}$ is the mutual escape speed for the colliding stars and $R_*$ is the axion star radius. To compute $v_{*,\mathrm{esc}}$, we follow \cite{Hertzberg:2020dbk} but allow $M_*$ and $R_*$ to be determined from our data set. Since $v$ will be determined by the halo environment, we can expect
\begin{align}
    \langle \sigma_* (v) v\rangle & = 4\pi\int p(v) \sigma_*(v) v^3 dv \; , \\
    p(v) & = p_0 \exp\left(-\frac{v^2}{2\sigma_v^2}\right) \; ,
\end{align}
where $\sigma_v$ is the velocity dispersion of the parent halo and $p_0$ is determined by 
\begin{equation}
    4 \pi \int_0^{v_\mathrm{esc}} p(v) v^2 dv = 1 \; ,
\end{equation}
where $v_\mathrm{esc}$ refers to the halo escape velocity.

To determine $\sigma_v$ we will make use of the stellar mass to velocity dispersion relation from~\cite{2016ApJ...832..203Z}
\begin{equation}
    \sigma_v = \sigma_b \left(\frac{M_\mathrm{stellar}}{M_b}\right)^\alpha \; ,
\end{equation}
with $M_b = 1.82 \times 10^{10}$ M$_\odot$, $\sigma_b = 118.3$ km s$^{-1}$, and 
\begin{equation}
    \alpha = \begin{cases}
        0.293 & M_\mathrm{stellar} > M_b \\
        0.403 & M_\mathrm{stellar} \leq M_b \\
    \end{cases} \; .
\end{equation}
Since our primary variable is the halo mass, we determine the stellar mass via the fitting function found in Equation~(3) of \cite{2013ApJ...770...57B}.

The total merger rate for axion stars within some cosmic volume $V$ can be determined via
\begin{equation}
    \Gamma_\mathrm{pop}(M_*) = \int \frac{d n}{d M_\mathrm{halo}} \Gamma_{**,\mathrm{halo}}  \ d M_\mathrm{halo} dV \; ,
\end{equation}
where $\frac{d n}{d M_\mathrm{halo}}$ is the halo mass function from \cite{Tinker_2008} as implemented in the \texttt{hmf} python package~\cite{Murray:2013qza}\footnote{We note that the updated package \texttt{TheHaloMod} \cite{Murray:2020dcd} from the same authors produces identical results for our purposes, as it includes the \texttt{hmf} code.}.



\section{Detectability}\label{sec:Detectability}

\begin{figure}[ht!]
    \centering
    \includegraphics[width=0.45\linewidth]{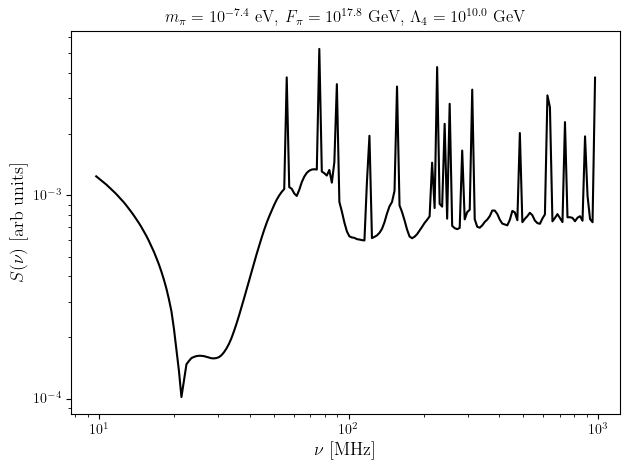}\includegraphics[width=0.45\linewidth]{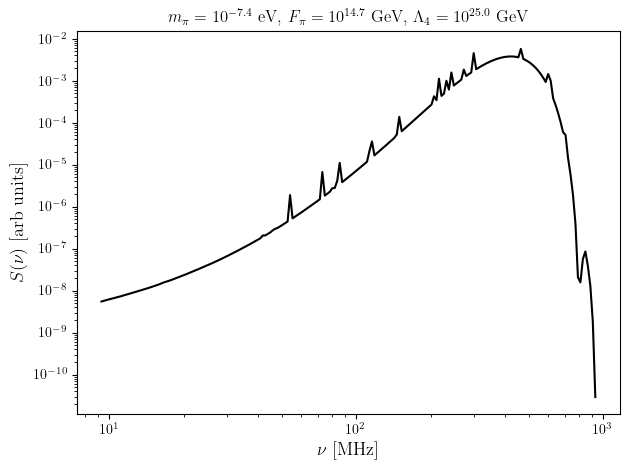}
    \includegraphics[width=0.45\linewidth]{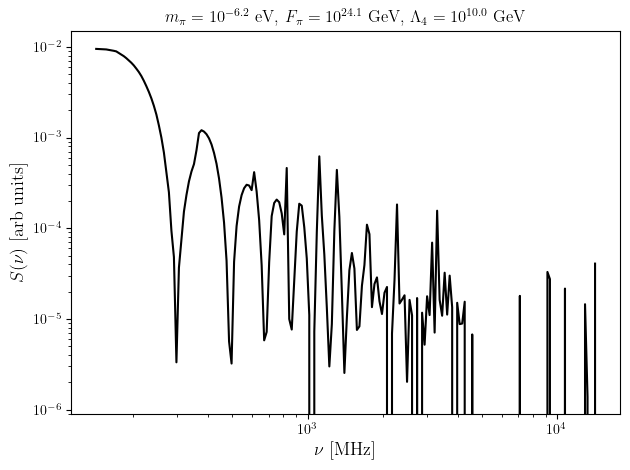}\includegraphics[width=0.45\linewidth]{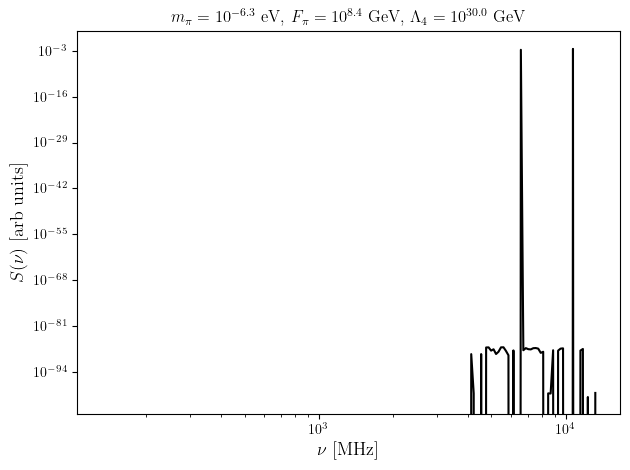}
    \includegraphics[width=0.45\linewidth]{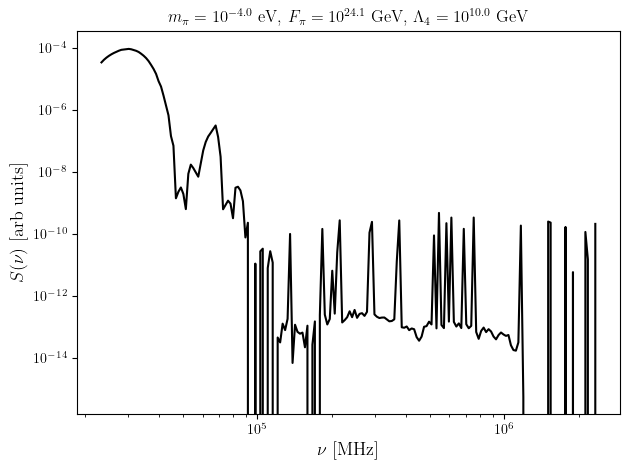}\includegraphics[width=0.45\linewidth]{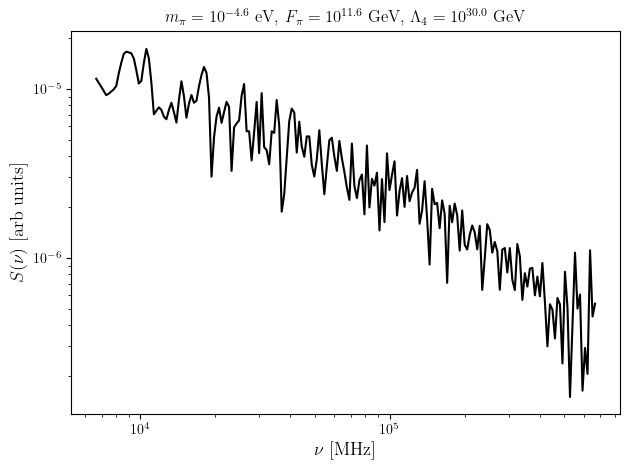}
    \caption{Example spectral shapes from Eq.~(\ref{eq:spec}). Each of these represents a broad category from the sampled parameter space. Spectra like those in the left column are associated with $\Lambda_4 \ll F_\pi$. In contrast, $F_\pi \ll \Lambda_4$ spectra are typified by the right-hand column. The frequency range is strongly dependent on $m_\pi$. The spectral shape has no apparent correlation with $\rho_\pi$, so that the determining factor between curves associated with given $F_\pi$ or $\Lambda_4$ ranges must be down to the local phases between the $\pi$-axion fields. }
    \label{fig:spectra}
\end{figure}

\begin{figure}[ht!]
    \centering
    \includegraphics[width=0.45\linewidth]{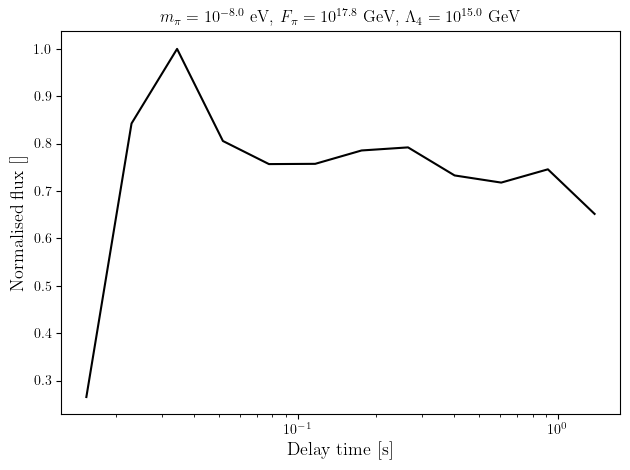}\includegraphics[width=0.45\linewidth]{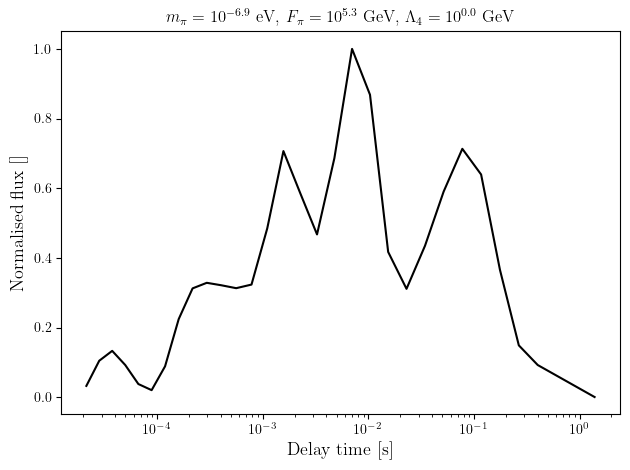}
    \includegraphics[width=0.45\linewidth]{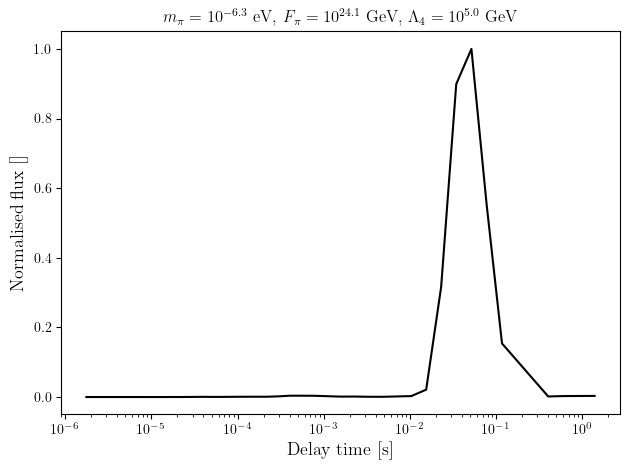}\includegraphics[width=0.45\linewidth]{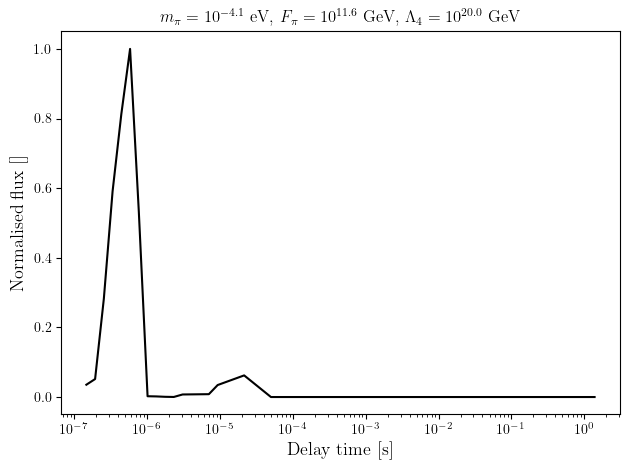}
    \caption{Example light curve shapes. These represent broad categories observed in the results. There is clearly a rich phenomenological structure in such multi-axion models. We note that the light curve strongly depends on $m_\pi$, but is also influenced by coupling strengths.}
    \label{fig:light-curves}
\end{figure}

Before showing our detectability results, we will display examples of the phenomenology of individual merger events. In particular, we display spectra in Fig.~\ref{fig:spectra}. In addition, we plot the light curves for a selection of events in Fig.~\ref{fig:light-curves}.   

\begin{figure}[ht!]
    \centering
    \includegraphics[width=0.45\linewidth]{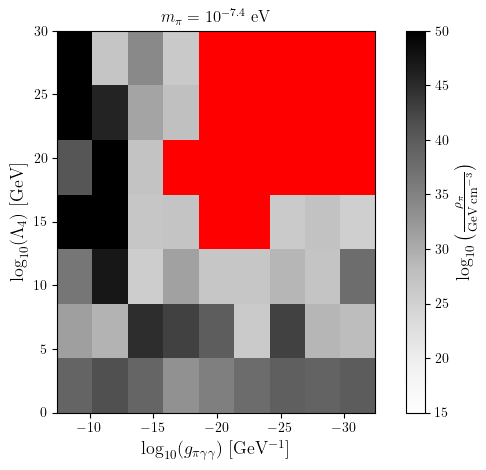}\includegraphics[width=0.45\linewidth]{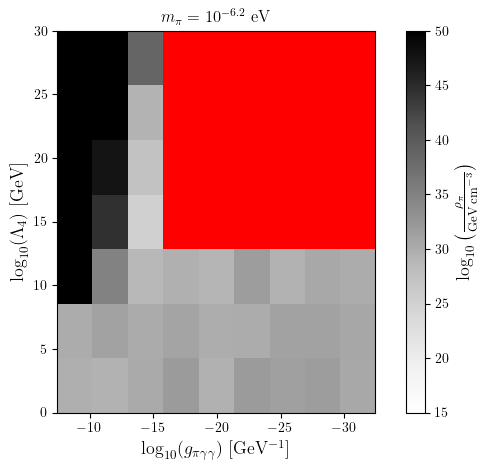}
    \includegraphics[width=0.45\linewidth]{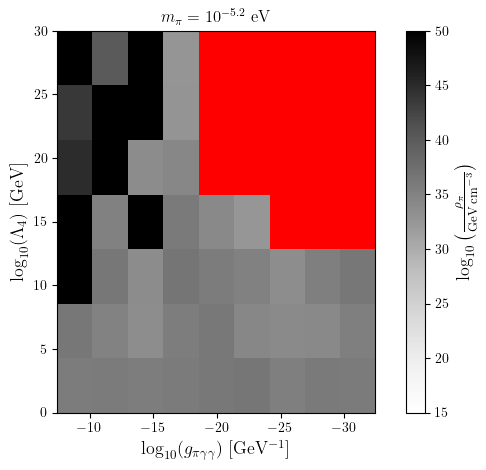}
    \caption{Required values of $\rho_\pi$ for $5\sigma$ transient observation with MeerKAT (flux in MeerKAT's band exceeds 50 $\mu$Jy). The luminosity distance used is $1$ Gpc. Note that $m_\pi$ is the mass of the lightest $\pi$-axion species. The red region indicates that where no observation is possible.}
    \label{fig:mk-limits}
\end{figure}

To display our results we consider a sample of $m_\pi$ values (this being the mass of the lightest species). For each we determine a flux as a function of $\rho_\pi$ (which we regard as the density of the merger remnant) and marginalize over all other parameters except $F_\pi$ and $\Lambda_4$. We then determine the $\rho_\pi$ required for observation of a merger associated with each point of the $F_\pi$-$\Lambda_4$ parameter space at the $5\sigma$ confidence interval. We display sample plots in Fig.~\ref{fig:mk-limits} for the MeerKAT telescope $550 \ \mathrm{MHz} \leq \nu \leq 1711 \ \mathrm{MHz}$ searching within 1 Gpc, where we assume the telescope observations have a typical RMS noise $\geq 10$ $\mu$Jy/beam to align with the order of magnitude of short duration transient searches in \cite{meertrap}. Here we can see that for masses $10^{-8}\ \mathrm{eV} < m_\pi < 10^{-5} \ \mathrm{eV}$, it is possible to have comparatively low density merger remnants that produce visible emissions ($10^{29} \ \mathrm{GeV}\ \mathrm{cm}^{-3} \leq  \rho_\pi \leq 10^{36} \ \mathrm{GeV}\ \mathrm{cm}^{-3}$). By ``low'', we are comparing to neutron star densities $\sim 10^{38}$ GeV cm$^{-3}$, which also tends to fall below the compact axion star threshold. Note that $\rho_\pi$ values above $10^{42}$ GeV cm$^{-3}$ are extrapolated from our data set and may not be reliable. Importantly, this density requirement need only apply to the merger remnant, so it does not commit us to having long-term stability for arbitrary densities. Typically, merger events are visible at low remnant densities if $g_{a\gamma\gamma} \leq 10^{-15}$ GeV$^{-1}$. Higher values of $m_\pi$ can mitigate this when the dilaton-like coupling is very strong, i.e. $\Lambda_4 \lesssim 10^{15}$ GeV.

\begin{figure}[h]
    \centering
    \includegraphics[width=0.45\linewidth]{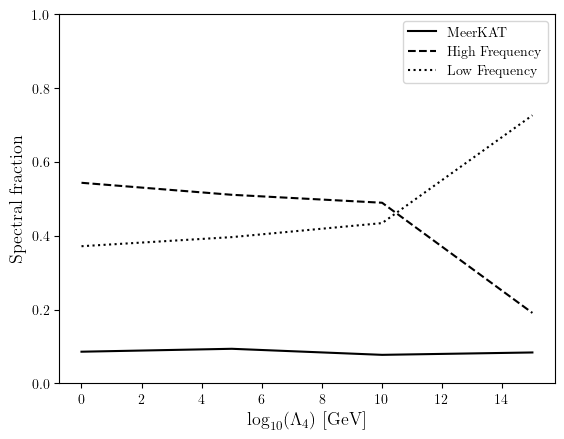}\includegraphics[width=0.45\linewidth]{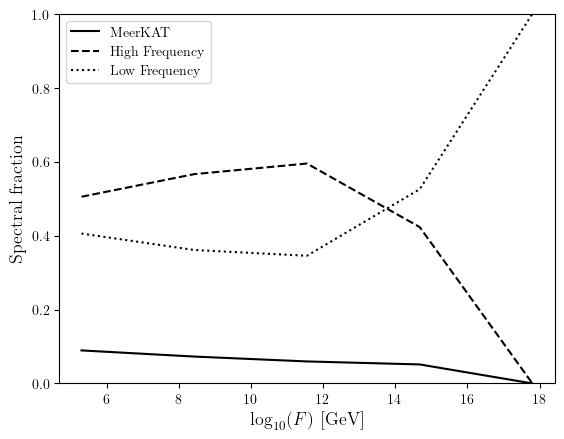}
    \includegraphics[width=0.45\linewidth]{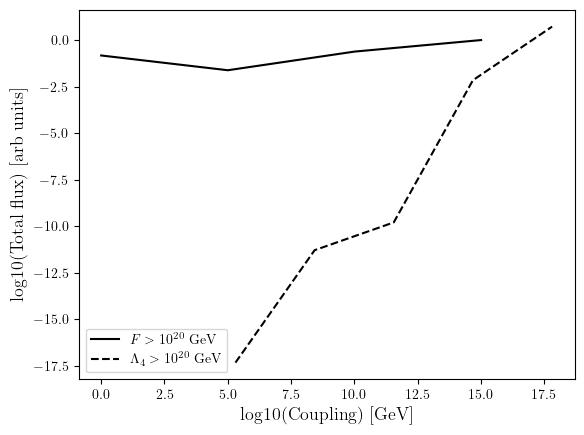}
    \caption{The effects of $F_\pi$ and $\Lambda_4$ couplings on spectral properties and total flux.}
    \label{fig:couplings}
\end{figure}

To fully understand these results we present Fig.~\ref{fig:couplings}, where we examine each coupling's effect independently. This is done by considering only cases where the other is $> 10^{20}$ GeV and marginalising over all other parameters. The upper panels reveals that the MeerKAT frequency band tends to contain around 10\% of the total flux. We also note that the upper panels of Fig.~\ref{fig:couplings} show that weak couplings favor emission frequencies below the MeerKAT band, whereas strong couplings prefer those above it. The lower panel shows the total flux can be suppressed by strong axion coupling when the dilaton-like coupling is weak. This is because stable axion stars in this region of parameter space have an extremely small size, due to their enhanced attractive self-interaction strength. Such tiny objects therefore experience very large rates of photon escape. In turn, this leads to a small mass threshold for stability and, consequently, very faint merger emissions that require boosting by very large densities to be visible. Notably, the dilaton-like coupling allows for a far greater range of stable stars. This implies it does not play an important role in determining the maximum allowed mass when the axion coupling is weak. However, it must be responsible for potentially observable mergers in the region where both couplings are strong. Additionally, the total flux is maximized when both couplings are weak. This allows for the largest stable stars, meaning the total radiated energy can compensate for the weaker couplings. If they become simultaneously too weak, then no resonance is achieved and mergers are too faint to observe.

For a look at future experiments we use the SKA and lower band of ngVLA, giving us a frequency range from $50 \ \mathrm{MHz} \leq \nu \leq 50 \ \mathrm{GHz}$, sensitive to transients fluxes  $\geq 1$ $\mu$Jy/beam. These results extend the parameter space we can constrain, particularly allowing us to probe masses down to $10^{-8}$ eV. We can cover a similar parameter space per mass, at 1 Gpc distance, as MeerKAT but only require $10^{27}\ \mathrm{GeV \, cm}^{-3} \leq \rho_\pi \leq 10^{34}\ \mathrm{GeV \,cm}^{-3}$ to do so for all masses $10^{-8}\ \mathrm{eV} < m_\pi < 10^{-5} \ \mathrm{eV}$, with higher masses requiring $\rho_\pi \sim 10^{40} \ \mathrm{GeV}\ \mathrm{cm}^{-3}$. The increased frequency range and sensitivity means we are less sensitive to the local phase configuration than for the MeerKAT instrument. Importantly, a pi-axion star merger is visible with axion couplings beyond even the reach of neutron star searches~\cite{Safdi:2018oeu,walters_axions_2024} and can target a wider mass range, thanks to the broader spectra that result from our model.

\begin{figure}[h]
    \centering
    \includegraphics[width=0.45\linewidth]{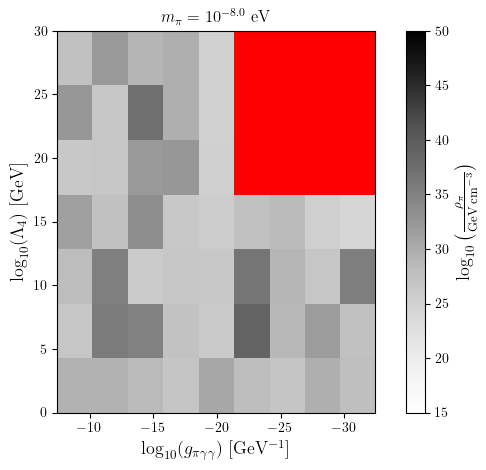}\includegraphics[width=0.45\linewidth]{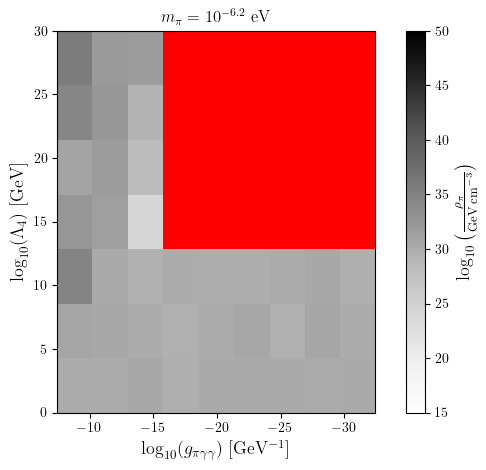}
    \includegraphics[width=0.45\linewidth]{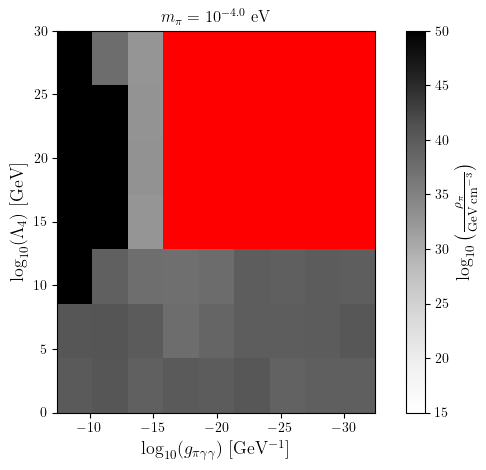}
    \caption{Required values of $\rho_\pi$ for $5\sigma$ transient observation with the SKA and ngVLA (flux in the band exceeds 5 $\mu$Jy). The luminosity distance used is $1$ Gpc. Note that $m_\pi$ is the mass of the lightest $\pi$-axion species. The red region indicates that where no observation is possible.}
    \label{fig:ska-limits}
\end{figure}

 \newpage

 \begin{figure}[h]
    \centering
    \includegraphics[width=0.45\linewidth]{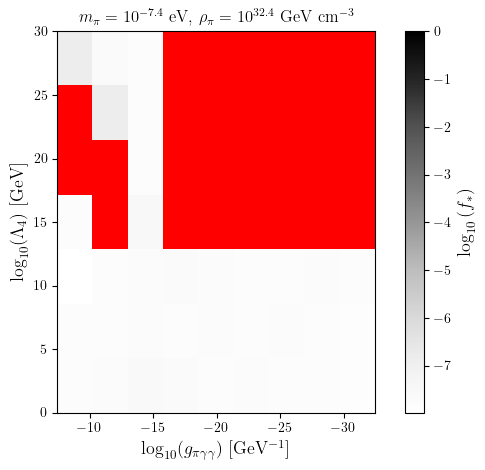}\includegraphics[width=0.45\linewidth]{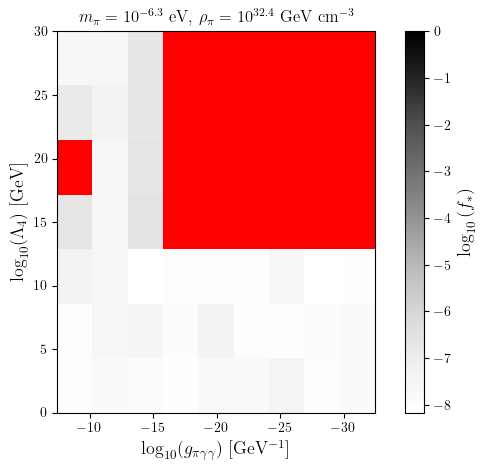}
    \includegraphics[width=0.45\linewidth]{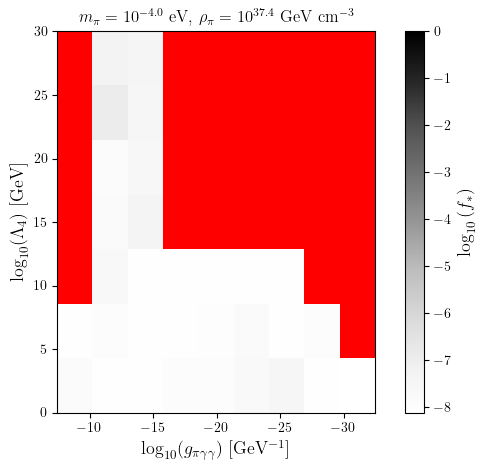}
    \caption{All-sky merger non-observation limits on the axion star population. The red region indicates no limits could be produced.}
    \label{fig:merger}
\end{figure}

\subsection{Population limits on $\pi$-axion stars}
 We will consider the number of mergers per year for a population within $z \leq 0.3$ and determine the limits that could be placed on the fraction of dark matter in the form of axion stars $f_*$ as a result of non-observation of a single merger event in a whole-sky survey. We display results for densities that can achieve $5\sigma$ detection thresholds at 1 Gpc distance in Fig.~\ref{fig:merger}. It is evident that such a survey could easily rule out a significant fraction of dark matter being in the form axion stars over a wide range of of the $m_\pi$-$F_\pi$-$\Lambda_4$ parameter space. However, we note that it is largely restricted to regions where at least one coupling is strong. Additionally, for $m_\pi = 10^{-8} \ \mathrm{eV}$, no constraints can be derived at densities that provide visible mergers. This is because low $m_\pi$ leads to larger star masses and smaller number densities.


\section{Conclusion}\label{sec:Conclusion}

An appealing feature of the $\pi$-axiverse is having a natural control on the separation of the ALPs' masses through $O(1-100)$ separation between the dark quark masses. Such physics is informed by the deeply understood and experimentally verified Standard Model. In this work, we have shown that the parametric resonance of photons interacting with the $\pi$-axiverse produces a unique electromagnetic spectra that depends on the $\pi$-axion masses and phases. We focused on a the case of $N_f=3$ and chose quark masses, and therefore $\pi$-axion masses, in a tight spectrum.  Sufficient resonance only occurs when the model couplings fall in certain regions.  As a  concrete physical system, we consider a homogeneous ($\pi$-)axion configuration above the critical threshold for resonance. The excess energy is radiated in part as photons\footnote{One can expect a gravitational wave (GW) signal as well, however the power radiated from the GWs is generally subdominant by a few orders of magnitude compared to the electromagnetic radiation. This is the case for single field axion models \cite{Chung-Jukko:2024hod}, however, we expect a similar approximation to hold in the $\pi$-axiverse provided the mass of the real scalar and axion coupling $F_\pi$ are of similar orders to the single field axion theory parameters.}  carrying specific spectral properties, which may then be detected and distinguishable from other astrophysical transient events of a similar nature, including single-species axion resonances. 

Our detectability results demonstrate that $\pi$-axion star mergers are visible to current and up-coming radio telescopes out to large cosmic distances for a large swath of model parameters, without requiring merger remnant densities above that of a neutron star. In particular, we can probe a much broader $\pi$-axion mass range than is possible in single-axion models. Additionally, we do not need to obey the single-axion requirement of $g_{a\gamma\gamma} F_a \geq 0.3$ to obtain parametric resonance in $\pi$-axion stars. These properties allow us to compensate for low per galaxy merger rates and probe a large region of the model parameter space through non-observation limits. We are able to probe axion star fractions as low as $10^{-8}$ with axion couplings $\geq 10^{-25}$ GeV$^{-1}$, as long as the dilaton-like coupling obeys $\Lambda_4 \lesssim 10^{15}$ GeV. We note, however, that this may be modified when accounting for an accurate spatial distribution of axion stars within dark matter halos~\cite{kavanagh_stellar_2021}. This demonstrates that this model is far easier to constrain than typical single-axion scenarios. For instance, mergers are observable with couplings below current neutron star constraints, or even SKA projections for these~\cite{Safdi:2018oeu,walters_axions_2024}.

Our analysis focused on the physics of an axion star-like system above a critical threshold undergoing spontaneous resonance.  It is straightforward to extend this method to include examples of stimulated decay, simply by modifying the photon field's initial conditions. Note that the initial conditions for the photon field do not modify the shape of the resonance spectrum, and have minimal effect on the timescale it takes for resonance to begin \cite{Chung-Jukko:2023cow}. One can consider photon emission from sub-critical $\pi$-axion stars triggered by incoming EM radiation from external astrophysical events. This could include incoming photons produced by other $\pi$-axion star mergers, but is also well-motivated when considering $\pi$-axion populations gathered around radio pulsars or within the disk of accreting black holes. For near-critical $\pi$-axion stars, a small perturbation in the photon field can push the system above its critical point, leading to parametric resonance and runaway photon production. However, in other regions of parameter space, or for sub-critical axion stars whose characteristic length scale is not large enough, there can be a finite but appreciable contribution to the photon field. Therefore, even in regions of parameter space where the $\pi$-axion stars may not be able to undergo runaway photon production and produce a detectable EM transient, they could still inject additional power into the spectra of transients produced from other astrophysical events. This can range from a flat power boost across all frequencies, to a change in the overall shape or peak frequency of the transient signal power spectrum. In contrast to examples of single-species axion resonant decay, the energy injected can be \textit{both} broad-band or narrow-band. We additionally note the time-dependent interference between $\pi$-axion species in configurations with more than one species of $\pi$-axion due to their incommensurate phases can produce highly non-trivial corrections to transient EM signals, leading to additional scintillation in the final signal power spectrum. Similar interactions may also produce a \textit{gegenschein}-like effect (\cite{Ghosh:2020hgd,Sun:2023gic}) with some of these additional spectral features, which would therefore be distinguishable from a spectrum produced by a single-axion configuration. Interactions between axions and photons can introduce additional polarization effects into emitted EM signals \cite{Chung-Jukko:2023cow}, dependent on the nature of the effective interaction vertices of the axionic theory with SM photons, as well as the symmetries in the geometric structure of the axion star \cite{Hertzberg:2018zte}. We leave a more robust analysis of such systems for future work.

The emergent complexity in the non-trivial interactions between $\pi$-axion species in the $\pi$-axiverse can result in energy injection and spectral shapes which are normally forbidden to emissions sourced by resonances with a single fundamental frequency, i.e. single-species axion decay. Our numerical searches of the physically permissible $\pi$-axiverse parameter space support the assertion that energy injection from $\pi$-axiverse resonant decay into standard model photons can come in the form of both narrow-band and broad-band transient signals. Additionally, unlike single-field models, the growth rate of photons does not follow a constant exponential trajectory but instead admits solutions which can enter and exit resonance repeatedly as time evolves. Therefore, regions of the standard $m_{a}$ - $g_{a\g\g}$ axion parameter space which are considered to be experimentally forbidden by experiments which rely on narrow-band resonances, e.g. resonant cavity searches and haloscope searches, are only partially constrained in the $\pi$-axiverse, even for the simplified $N_f=3$ case, and before considering the additional resonance channel available through the $FF$ term of the $\pi$-axiverse field equations, in Eq. \eqref{LU(1)}. Furthermore, we find that stimulated decay from $\pi$-axiverse populations can contribute appreciable but sub-dominant additional power into the spectrum of propagating EM signals sourced by background astrophysical events. This additional power can take the form of a smooth increase in power across all frequencies, leading to an effective shift in the peak frequency of the final spectrum, or even a spectrum with a visible secondary peak. Alternatively, rapid oscillations in the $\pi$-axion and photon amplitudes caused by interferences between species across different frequencies can also emerge in the final power spectrum as additional scintillation. Therefore, future searches which hope to use astrophysical transient data to constrain parameter spaces of theories which admit multiple species of ALPs should consider these additional potential contributions to the final spectrum when using it to estimate source parameters or perform noise characterization. We note that this behavior could also leave detectable imprints on measurements of other relevant classes of transient events beyond those considered in detail in this analysis, such as gravitational wave (GW) surveys of ultralight scalar bosonic DM \cite{Cole:2022yzw,Alexander:2018qzg,Wilcox:2024sqs,Xie:2024ubm}. While the widely unconstrained $\pi$-axiverse parameter space certainly permits solutions which are in regions of interest for such surveys, we leave a more detailed investigation to future work.

We restricted to the subregion of parameter space where $N_f=3$ and $\varepsilon=1$ for simplicity and computational practicality. This effectively resulted in three electrically neutral states in the relic spectrum. Studying a larger value of $N_f$ will be informative of how the physics changes as you add more states, and thus more driving frequencies, to the resonant structure. Qualitatively, for $N_f\gg 1$, we may expect the resonance structure to commonly exhibit a broader spectrum do to a larger amount of driving frequencies stemming. We also may find an increased dependence of resonance to the relative phases of the coherent $\pi$-axions. Unfortunately, due to the complexity of (\ref{modeEOM}) and the lack of closed-form solutions or valid approximations of their asymptotic behavior, in practice we are restricted to brute-force numerical approaches, which are in turn limited by computational power. Even so, studying the case of $N_f=4$ gives us a total of 15 states, which may considerably change the results in this paper. If the charged states are superheavy and not part of the relic spectrum, we then have three real $\pi$-axions and two electrically neutral, complex $\pi$-axions (see (\ref{evenNstates})), which may be computationally feasible and quantitatively different from the $N_f=3$ case. Allowing $\varepsilon<1$ also may change the resonant behavior, and can move the $g_{\pi\g\g}$ coupling away from the constrained regions \cite{Alexander:2023wgk} particularly for masses in the $10^{-3}$ eV range. Finally, we focused our analysis on the $\pi$-axion masses lying in a tight spectrum separated by $O(1)$ factors. By sampling the $c_i$ constants across a larger range of values (see Table \ref{tab:ci_table}), we can explore the effects of a wider mass spectrum on the resonant structure. In summary, different regions of the multidimensional parameter space merit further investigation.

\subsection*{Acknowledgments}
The authors thank Humberto Gilmer, Luca Visinelli, and Hong-Yi Zhang for helpful discussions and comments.  S.A. and T.M. acknowledge support from the Simons Foundation, Award 896696. G.B. acknowledges support from The University of the Witwatersrand and Brown University in funding a visiting researcher program. Part of this research was conducted using computational resources and services at the Center for Computation and Visualization, Brown University. This work also made use of the Illinois Campus Cluster, a computing resource that is operated by the Illinois Campus Cluster Program (ICCP) in conjunction with the National Center for Supercomputing Applications (NCSA) and which is supported by funds from the University of Illinois Urbana-Champaign.

\newpage

\bibliography{refs}
\end{document}